\RequirePackage{fix-cm}
\documentclass{svjour3}                     
\smartqed  
\usepackage[a4paper, total={6in, 9.5in}]{geometry}
\usepackage{bm}
\usepackage{latexsym}
\usepackage{dcolumn}
\usepackage{amsfonts,amssymb}
\usepackage{graphicx}
\usepackage{epsfig}
\usepackage{psfrag}
\usepackage{amsmath}
\usepackage{enumerate}
\usepackage[hang,flushmargin]{footmisc}
\usepackage[titletoc,toc]{appendix}
\usepackage{multirow}
\usepackage{lipsum}
\usepackage{epstopdf} 
\usepackage{hyperref}
\usepackage{rotating}
\usepackage{multirow}
\usepackage{subcaption}
\newcommand{\be}{\begin{equation}}
\newcommand{\ee}{\end{equation}}
\newcommand{\bea}{\begin{eqnarray}}
\newcommand{\eea}{\end{eqnarray}}
\newcommand{\bse}{\begin{subequations}}
\newcommand{\ese}{\end{subequations}}
\newcommand{\bce}{\begin{center}}
\newcommand{\ece}{\end{center}}
\newcommand{\bfg }{\begin{figure}}
\newcommand{\efg}{\end{figure}}
\newcommand{\bi}{\begin{itemize}}
\newcommand{\ei}{\end{itemize}}
\newcommand{\bed}{\begin{description}}
\newcommand{\eed}{\end{description}}
\newcommand{\ben}{\begin{enumerate}}
\newcommand{\een}{\end{enumerate}}
\newcommand{\nn} {\nonumber}

\newcommand{\la}{\label}
\def \le {\left}
\newcommand{\ri}{\right}
\newcommand{\pa}{\partial}
\newcommand{\fr}{\frac}
\newcommand{\sq}{\sqrt}

\newcommand{\lra}{\longrightarrow}

\def\a  {\alpha}
\def\b  {\beta}
\def\c  {\gamma}
\def\C  {\Gamma}
\def\d  {\delta}

\def\e  {\epsilon}
\def\E  {\eta}
\def\f  {\phi}
\def\F  {\Phi}
\def\k  {\kappa}
\def\l  {\lambda}
\def\L  {\Lambda}
\def\m  {\mu}
\def\n  {\nu}
\def\o  {\omega}
\def\O  {\Omega}

\def\r  {\rho}

\def\s  {\sigma}

\def\vph {\varphi}

\newcommand{\cA}{\mathcal A}

\newcommand{\cF}{\mathcal F}

\newcommand{\cT}{\mathcal T}

\newcommand{\cL}{\mathcal L}

\newcommand{\cQ}{\mathcal Q}
\newcommand{\cR}{\mathcal R}
\newcommand{\cM}{\mathcal M}
\newcommand{\cN}{\mathcal N}
\newcommand{\cS}{\mathcal S}
\newcommand{\cV}{\mathcal V}

\newcommand{\rX}{\r_{\!_{X}}}
\newcommand{\rmf}{\r^{(m)}_{\text{\small eff}}}
\newcommand{\Omf}{\O^{(m)}_{\text{\small eff}}}
\newcommand{\OX}{\O_{\!_X}}
\newcommand{\oX}{\o_{\!_X}}
\def\Fp {\F_{_0}}
\def\cUp {\cU_{_0}}
\def\cVp {\cV_{_0}}

\def\fp {\f_{_0}}
\def\tp {t_{_0}}
\def\Geff {G_{\text{\small eff}}}

\newcommand{\cU}{\mathcal U}

\newcommand{\fw}{\mathfrak w}
\newcommand{\Rt}{\widetilde{\cR}}
\newcommand{\Ct}{\widetilde{\Gamma}}
\newcommand{\nab}{\nabla\!}
\newcommand{\nt}{\widetilde{\nabla}\!}

\newcommand{\hg}{\hat{g}}
\newcommand{\hR}{\hat{\cR}}
\newcommand{\hS}{\hat{\cS}}
\newcommand{\hLm}{\hat{\cL}_{m}}

\newcommand{\bdm}{\begin{displaymath}}
\newcommand{\edm}{\end{displaymath}}

\long\def\symbolfootnote[#1]#2{\begingroup%
\def\thefootnote{\fnsymbol{footnote}}\footnote[#1]{#2}\endgroup}
\numberwithin{equation}{section}

\begin{document}

\title{Phase Plane Analysis of Metric-Scalar Torsion Model for Interacting Dark Energy
}


\author{Arshdeep Singh Bhatia         \and
        Sourav Sur 
}


\institute{Arshdeep Singh Bhatia \at
              Department of Physics \& Astrophysics \\ University of Delhi, New Delhi-110007 \\
              \email{arshdeepsb@gmail.com, asbhatia@physics.du.ac.in}          
           \and
           Sourav Sur\at
               Department of Physics \& Astrophysics \\ University of Delhi, New Delhi-110007 \\
              \email{sourav.sur@gmail.com, sourav@physics.du.ac.in}          
}

\date{Received: date / Accepted: date}

\maketitle

\begin{abstract}
We study the phase space dynamics of the non-minimally coupled Metric-Scalar-Torsion model in both \emph{Jordan} and  \emph{Einstein frames}.  
We specifically check for the existence of critical points which yield stable solutions 
representing the current state of accelerated expansion of the universe fuelled by the \textit{Dark Energy}. It is found that such solutions do 
indeed exist, subject to constraints on the
free model parameter. In fact the evolution of the universe at these stable critical points exactly
matches the evolution given by the cosmological solutions we found analytically in our previous work on the subject. 
\end{abstract}

\section{Introduction}  \label{sec:intro}

Among the most pertinent questions in science today is that of the origin and evolution of the \emph{dark energy} (DE). 
The discovery of the late time acceleration of the universe has led to a belief in existence of a repulsive gravity 
inducing DE component which amounts for nearly three 
quarters of the total energy content of our universe \cite{planck}. Although the \textit{cosmological constant} ($\L$) does 
a reasonably good job of explaining the observations, the underlying roots and composition of thus obtained DE sector are still a 
mystery. Besides the cosmological constant approach suffers from some 
deep theoretical issues, the \textit{fine tuning} and the \textit{coincidence} problems being the most prominent 
ones \cite{lambda}. This has led to the development of a plethora of models where a known source of energy-momentum 
tensor, say the scalar field, mimics the role of DE. Owing to the homogeneous and isotropic nature of the universe on 
large scales, scalar fields are infact one of the prime contenders for the role of the DE. Existence of scalar fields in a cosmological framework is not a new idea and has been around for decades, with works such as the \textit{Brans-Dicke} theory appearing as early as $1960s$ \cite{bransd}.

Some of the more popular scalar field models are the quintessence, K-essence, chaplygin gas etc. \cite{sami,sclrmodels}. But these models mostly lie in the regime of minimal coupling between the scalar field and gravity. A more general class of theories also include an explicit coupling between the
scalar filed and some geometrical artefact constructed out of the metric tensor. The most relevant example of such a theory is the aforementioned Brans-Dicke theory where the scalar field couples non-minimally to the Ricci scalar, $\cR$. These theories however fail to provide any physical (or even mathematical for that matter) motivation behind including such non-minimally coupled terms. One can also modify the geometrical constructs 
of the space-time appearing in the lagrangian.  The $f(R)$ models and inclusion of \textit{Gauss-Bonnet} terms are a couple of such attempts to explain the universe as we see it by going beyond the usual Hilbert action for the gravity sector \cite{fR}. Either way, all these models either
introduce a new source of energy-momentum tensor (via the scalar field in one form or the other) or aim to modify the terms representing the space-time geometry in a usual formulation of General Relativity (GR). \footnote{Of late there has been an increasing impetus to move away from this paradigm and to interpret a given cosmology in light of other fundamental interactions. See for example \cite{cruz}. }

But how about extending instead the definition of geometry itself? One such generalization of GR is constructed by lifting the assumption of having symmetric affine connections in the theory - the Christoffels, and replacing them with a more general class of asymmetric (but metric compatible) affine connections. A two index antisymmetriztion of these new connections leads to introduction of a third rank tensor field called
\textit{Torsion}\cite{hehl,akr,traut,sab,shap,ssasb}. Torsion, which geometrically manifests the classical spin density was first proposed to exist in early 1920s by Einstein and Cartan. In modern theories, torsion provides a classical background for quantized matter with spin and as such could be a low energy manifestation of a quantum gravitational theory \cite{shap,ham}. The massless antisymmetric Kalb-Ramond field in a closed string theory for
example, can source a completely antisymmetric torsion in a low energy limit \cite{pmssg,saa}. Physical implications of such an induced torsion is studied in considerable detail in \cite{ssgss,skpm,dmpmssg,acpm,bmssgsen,ssgss1,bmssgss}. The $f(T)$ models in teleparallel theories \cite{fT} and the Poincar\'e gauge theory of gravity \cite{yonest} are some notable theories which have treated torsion in a cosmological setup.  A classic compilation of works on classical gauge theories of gravity may be found in \cite{hehlrev}.

We proceeded along these lines in \cite{ssasb2} to developed a model of universe which has matter and a scalar field as it's residents, all in presence of torsion. In this setup, we were able to motivate a non-minimal coupling between the scalar field and gravity sectors. Working in the
standard cosmological setup, the nature of lagrangian which emerged for this model allowed us to completely trade away the torsion field in favour of augmented scalar field terms. It was shown that the model effectively yielded the Brans-Dicke action in the Jordan frame. In Einstein frame, the model
reduced itself to a quintessence model but with an additional non-minimal coupling between the scalar field and matter sectors. Most importantly we were able to analytically solve the resulting field equations (a set of coupled differential equations) for the evolution of the universe and it's components in both Jordan and Einstein frames.

Obtaining solutions however for a set of coupled differential equations is just half the story. Would these solutions 
persist over time once subjected to fluctuations in the phase space? Which is to say: are these solutions stable? It is this question that we answer in this work. To this end, we reduce the field equations found in \cite{ssasb2} to a set of autonomous equations whose equilibrium 
solutions, known as critical points, represent the state of the system (our universe in this case) in an asymptotic limit. The stability of the system at each critical point can be analysed by employing some straightforward mathematical recipes. In a case 
where multiple critical points co-exist for a system, one needs to analyse the dynamics of the system at each of those points and figure out the ones which paint a meaningful picture of the system. Such an analysis for the pure quintessence model 
has been done and studied extensively in literature \cite{sami}. The existence of cosmological scaling solutions in 
presence of exponential potentials has also been analysed in detail in \cite{copelandliddle}. Analysis in case of non exponential potentials may be found in \cite{sprt}. The phase space analyses of tachyon fields and dilatonic ghost condensates is discussed in \cite{sami},\cite{huang}. Analysis of some other forms of 
interactions between the dark energy and matter sectors is dealt in \cite{chengong}. 

The stability analysis for the model under consideration is performed in both the Jordan and the Einstein frames. Apart from ascertaining the stability, we also look into the dynamics of the system (and it's constituents) leading upto and at the critical points. To this end, we manage to
extract the evolution some cosmological parameters of interest, like the effective dark energy density parameter ($\OX$) and equation of state ($\oX$) over time. These parameters provide a useful insight into the state the system is in at any given point of time. Over the course of this work, we
employ both analytical and numerical techniques to achieve our objectives. We present the Einstein frame results first as the mathematical formulation in this frame is consistent with the canonical formulation of GR. In this frame, although we setup our system exactly in line with the corresponding analysis for a pure quintessence field (in terms of the matter and scalar field sectors), we recast and study the results in terms of an effective
matter and DE sectors instead. This is done to facilitate a meaningful correspondence with cosmological observations. The scalar field and matter sectors as we know remain coupled to each other in the Einstein frame. Hence they interact mutually and effect the overall dynamics of each other. In such a scenario, it becomes meaningless to talk about either sector without invoking the other, rendering any comparison with observed data redundant.
It is ensured while defining the effective matter and DE sectors that they remain mutually decoupled at all times. In terms of these decoupled
components, we find some very non-trivial results not seen in the corresponding quintessence analysis. For instance, physical compatibility of the results in this frame dictates an eventual extinction of the matter sector, irrespective of the initial conditions. Also, subjecting the theoretically
obtained value of the cosmological parameters to their physical bounds allows us to put constraints on the free model parameter. Additionally, the evolution and the overall dynamics of the system is interpreted in terms of the irreducible torsion components. This is made possible due to the fact
that the trace ($\cT_{\m}$) and the pseudo-trace ($\cA_{\m}$) modes of torsion, in context of the model under scrutiny, can directly be related to the kinetic and the potential energy terms of the scalar field respectively. Contribution of either torsion mode towards the overall energy density of the system and their relative strengths are also obtained over the course of this work.

The above analysis is then repeated for the Jordan frame formulation of the model. Although in this frame there is no explicit non-minimal coupling between the matter and scalar field sectors, the fact that the coupling constants are now time dependant gives rise to some interesting consequences. In a very succinct and concise manner, we investigate these effects and of course the overall stability of the Jordan frame solutions. It must be mentioned that although this work is presented as a follow up to our earlier work, it has the potential to stand as an independent work in the field of scalar-tensor theory.

The organization of this paper is as following: In section \ref{sec:mst} we briefly recapitulate the salient 
features of the Metric-Scalar-Torsion (MST) model introduced in \cite{ssasb2}. We then in section \ref{sec:field_eqns} summarize the field equations governing the evolution of the universe and it's constituent components as obtained in the Einstein frame.  From these field equations 
we then in  section \ref{sec:phase plane analysis} construct an autonomous system of differential equations which represent our original model in a phase space. Systematic analysis of these equations is undertaken for possible solutions and their inherent stability. We discuss the resulting cosmologies at each of the existing critical points and single out the one point
relevant to us. In section \ref{sec:numeric}, we scrutinize our model using some numerical recipes for solving a set of coupled differential equations. Finally in section \ref{sec:jordan}, we analyse the Jordan frame formulation of the MST model for possible equilibrium solutions and their inherent stability. In section \ref{sec:conclusion} we conclude this work by summarizing our main results and some possible extensions to this work. The appendix gives a heuristic account of the underlying mathematical theory of autonomous systems of differential equations employed in this work.

The notations and conventions we follow in this work are as following: the metric 
signature throughout this paper is $ (-,+,+,+)$. We work in units where speed of light, $c$, is taken to be unity. The factor $\k^2 \equiv 8 \pi G$ where $G$ is the Newton's gravitational 
constant.

\section{The MST model}  \la{sec:mst}

Let us begin this work by first presenting a brief overview of the Metric Scalar Torsion (MST) theory. A four dimensional Riemann-Cartan space-time ($U_4$) is a generalization of the four dimensional Riemannian space-time ($R_4$) in presence of torsion. The $U_4$ space-time is 
characterized by an asymmetric affine connection: $\Ct^{\l}_{~~\m\n} (\neq \Ct^{\l}_{~~\n\m})$. This connection is related
to the torsion tensor ($ T^{\l}_{~~\m\n}$) by the following definition:

\be
T^{\l}_{~\m\n} ~:=~ \Ct^{\l}_{~\m\n} - \Ct^{\l}_{~\n\m}
\ee
The torsion tensor can decomposed into three irreducible modes as:

\be
T^{\l}_{~\m\n} ~=~ \fr 1 3 \le(\d^{\l}_{\n} T_{\m} - \d^{\l}_{\m} T_{\n} \ri) + \fr 1 6 \e^{\l}_{~\m\n\s} \cA^{\s} + Q^{\l}_{~\m\n}
\ee
where $T_{\m} := T^{\n}_{~\m\n}$ is the torsion trace mode, $\cA^{\s} := \e^{\a\b\c\s} T_{\a\b\c}$ is the torsion pseudo-trace mode
and $Q_{\m\n\s}$ is the (pseudo-)tracefree mode of torsion.
In presence of torsion, the $R_4$ covariant derivative ($\nab_{\m}$) defined in terms of the Riemannian Levi-Cevita connections, 
$\C^{\l}_{~\m\n}(= \C^{\l}_{~(\m\n)})$ is generalized to to it's $U_4$ analogue, ($\nt_{\m}$), defined in terms of $\Ct^{\l}_{~\m\n}$ such that the metricity condition, $\nt_{\m} g_{\a\b} = 0$ is always satisfied. The curvature scalar in a $U_4$ space-time can similarly be expressed as:

\be
\Rt ~=~ \cR + \Gamma
\ee
where $\cR$ is the $R_4$ curvature scalar while $\cT$ is the part made up of torsion and expressed as:

\be
\Gamma ~=~ -2\nab_{\m} T^{\m} - \fr 2 3 T^{\m} T_{\m} + \fr 1 {24} \cA^{\m} \cA_{\m} + \fr 1 2 Q_{\m\n\s} Q^{\m\n\s}
\ee
For a detail account of formalism in presence of torsion refer \cite{ssasb}. 

As discussed in detail in \cite{ssasb2}, the canonical scalar field action in a $U_4$ space-time suffers from an inherent ambiguity. 
Unlike their counterparts in $M_4$ (four dimensional Minkowski) and $R_4$ space-times, the $U_4$ lagrangians of scalar field ($\f$) 
:

\be
\cL_(\f) ~=~ - \fr 1 2 g^{\m\n} \nt_{\m} \f \nt_{\n} \f -V(\f)
\ee
and
\be
\cL^{'}_(\f) ~=~  \fr 1 2 g^{\m\n} \f ~ \nt_{\m}  \nt_{\n} \f -V(\f)
\ee
do not differ by a total divergence term and are hence not equivalent. Here the self interacting potential of the scalar field was denoted by $V(\f)$. One is thus met by a dilemma as to which lagrangian to start with while constructing a scalar field action in a MST theory. The resolution of this problem may lie in generalizing the standard Gibbons-Hawking-York boundary term in presence of torsion and the scalar field. As an easier alternative to this, we proposed having a counter-term in the action which would ensure the difference between the two lagrangians is at most a total divergence term. Such a term can be obtained 
by non-minimally coupling the scalar field to the trace mode of torsion. A more elegant way of doing this is to couple the $U_4$ curvature scalar ($\Rt$) to the scalar field non-minimally. This ensures a coupling between the scalar field and the trace mode of torsion. The MST action hence, upto a divergence term can be written as:

\bea \la{MST-ac}
\cS \,=\, \int d^4 x \, \sq{-g} \le[\fr{\b \f^2} 2 \le(\cR \,+\, 4 \, \cT^\m \, 
\fr{\pa_\m \f}{\f} \,-\, \fr 2 3 \, \cT_\m \cT^\m \,+\, \fr 1 {24} \, \cA_\m \cA^\m 
\,+\, \fr 1 2 \cQ_{\m\n\s} \cQ^{\m\n\s}\ri) \ri. \nn\\
\le. -\, \fr 1 2 \, g^{\m\n} \, \pa_\m \f \, \pa_\n \f \,-\, 
V (\f) \,+\, \cL^{(m)} \ri] \,,
\eea
Here $\b$ is a positive coupling constant and $\cL^{(m)}$ denotes the presence of matter fields. The trace mode of torsion ($\cT_\m$) appears as an auxiliary field in the above action. Variation of (\ref{MST-ac}) with respect to it gives:

\be \la{tr-eqn}
\cT_\m ~=~ 3 \fr{\pa_\m \f}{\f}
\ee
We hence find the scalar field sourcing one of the torsion modes. Since torsion is conventionally taken to be massless, we would expect it's source field to be massless as well. However we allow the scalar field to have a mass $m_\f$ resulting in a pontentials of type $V(\f)=m_\f ^2 \f^2 / 2$. Additionally we consider only a very slow rolling (effectively a constant) pseudo-trace mode of torsion, $\langle \cA_\m \rangle $. The homogeneous and isotropic nature of the observed universe prompts us to use the Friedmann-Robertson-Walker (FRW) setup but this in turn puts severe constraints on the existence of the torsion modes. For an in-depth analysis of these constrants, see \cite{ssasb}. In view for these constraints and Eq. (\ref{tr-eqn}), the torsion contribution to the system can effectively be written in terms of augmented scalar field terms. Accounting for these inputs, the action now looks like:
\be \la{MST-ac2}
\cS \,=\, \int d^4 x \, \sq{-g} \le[\fr{\b \f^2} 2 \, \cR \,-\, \fr{\le(1 - 6\b\ri)} 2 
\, g^{\m\n} \, \pa_\m \f \, \pa_\n \f \,-\,\fr 1 2 \, m^2 \f^2 \,+\, \cL^{(m)} \ri] \,.
\ee
with
\be
\fr 1 {2} m^2 \f^2  ~=~ V(\f) + \fr {\b \f^2} {48} \langle \cA \rangle^2
\ee
This is of course the action of a scalar-tensor theory in the original {\em Jordan} 
frame \cite{frni}, which is characterized by a running gravitational coupling parameter
\be \la{G-eff}
\Geff (t) \,=\, G \le[\fr{\fp}{\f (t)}\ri]^2 \,, 
\ee
defined so that at the present epoch $\, t = \tp$, we have $\, \Geff (\tp) = G$, the 
Newton's constant, under the stipulation
\be \la{phi0}
\f (\tp) \, \equiv \, \fp \,=\, \fr 1 {\k \sq{\b}} \,\,, \qquad \mbox{with} \qquad 
\k = \sq{8 \pi G} \,\,.
\ee
If one redefines the scalar field as
\be \la{J-phi}
\F (t) \,:=\, \Fp \le[\fr{\f (t)} \fp\ri]^2 \,\,, \qquad \mbox{with} \qquad
\Fp \,\equiv\, \F (\tp) \,=\, \b \, \fp^2 \,=\, \k^{-2} \,\,, 
\ee
then Eq. (\ref{MST-ac2}) reduces to an equivalent {\em Brans-Dicke} (BD) action (albeit 
with a potential for $\F$) \cite{frni}:
\be \la{J-ac}
\cS_J \,=\, \int d^4 x \, \sq{-g} \le[\fr{\F \, \cR} 2 \,-\, \fr \fw {2 \F} \, g^{\m\n} \, 
\pa_\m \F \, \pa_\n \F \,-\, \cV(\F) \,+\, \cL^{(m)} \ri] \,,
\ee
where $\fw$ is the effective BD parameter, given by
\be \la{BD-param}
\fw \,=\, \fr 1 {4 \b} \,-\, \fr 3 2 \,\,,  
\ee
and the scalar field potential is 
\be \la{J-pot}
\cV (\F) \,=\, \cVp \, \fr \F \Fp \,\,, \qquad \mbox{with} \qquad
\cVp \,\equiv\, \cV (\F) \big|_{t=\tp} =\, \fr 1 2 \, m^2 \fp^2 \,\,.
\ee
Under a conformal transformation of type, 
\be \la{conf}
g_{\m\n} \,\lra\, \hg_{\m\n} \,= \le(\fr \f \fp\ri)^2 g_{\m\n} \,\,,
\ee 
Eq. (\ref{MST-ac2}) reduces to the scalar-tensor action in the so-called {\em Einstein} frame:
\be \la{E-ac}
\hS \,=\, \int d^4 x \, \sq{-\hg} \le[\fr{\hR}{2 \k^2} \,-\, \fr 1 2 \le(\fr \fp \f\ri)^2
\hg^{\m\n} \, \pa_\m \f \, \pa_\n \f \,-\, \fr {m^2 \fp^2} 2 \le(\fr \fp \f\ri)^2 \,+ 
\le(\fr \fp \f\ri)^4 \cL^{(m)} \!\le(\hg,\f\ri) \ri] \,,
\ee
where $\, \hg \equiv \text{det} \le(\hg_{\m\n}\ri)$, and $\, \hR = \hg^{\m\n} \hR_{\m\n}$ 
with $\, \hR_{\m\n}$ the Ricci tensor constructed using $\, \hg_{\m\n}$. Note that 
$\, \cL^{(m)}$ is now in general a function of both $\hg_{\m\n}$ and $\f$. Redefining, 
without loss of generality, the scalar field as
\be \la{E-phi}
\vph \,:=\, \fp \, \ln \le(\fr \f \fp\ri) \,\,, \qquad \mbox{such that} \qquad
\vph (\tp) = 0 \,\,,
\ee
the Einstein frame action (\ref{E-ac}) can be expressed in a more convenient form:
\be \la{E-ac1}
\cS_E \,=\, \int d^4 x \, \sq{-\hg} \le[\fr{\hR}{2 \k^2} \,-\, \fr 1 2 \, \hg^{\m\n} \, 
\pa_\m \vph \, \pa_\n \vph \,-\, \cU (\vph) \,+\, \hLm \!\le(\hg,\vph\ri) \ri] \,,
\ee
with the corresponding matter Lagrangian density 
\be \la{E-mat}
\hLm \!\le(\hg,\vph\ri) \,=\, e^{- 4 \vph/\fp} \, \cL^{(m)} \!\le(\hg,\f (\vph)\ri)
=\, e^{- 2 \k \sq{\b} \vph} \, \cL^{(m)} \!\le(\hg,\f (\vph)\ri) \,,
\ee
and the scalar field potential given as
\be \la{E-pot}
\cU (\vph) \,=\, \cUp \, e^{- 2 \vph/\fp} \,=\, \cUp \, e^{- 2 \k \sq{\b} \vph} \,\,,
\qquad \mbox{where} \qquad \cUp \,\equiv\, \cU (\vph) \big|_{t=\tp} =\, \fr{m^2 \fp^2} 2 
\,=\, \fr{m^2}{2 \k^2 \b} \,\,.
\ee

As shown above, the torsion components and the scalar field are inter-related. This facilitates interpreting a system and it's properties in terms of variables belonging to either of these sectors. One such set of variables are the norms of the torsion components, $|\cT|$ and $|\cA|$. The former is  related to the scalar field in the Jordan frame as:

\be \label{norm_T_jordan}
|\cT| ~:=~ \sq{-g^{\m\n}\cT_\m \cT_\n} ~=~ \fr 3 {2 \F} \sq{-g^{\m\n}\pa_\m \F \pa_\n \F}
\ee
whereas in the Einstein frame,

\be \label{norm_T_einstein}
|\cT| ~:=~ \le( \fr{\f}{\fp} \ri) \sq{-\hat{g}^{\m\n}\cT_\m \cT_\n} ~=~ 3 \k \sq{\b} e^{\k \sq{\b} \vph} \sq{-\hat{g}^{\m\n}\pa_\m \vph \pa_\n \vph}
\ee
Norm of the pseudo-scalar part in both Jordan and Einstein frames evolves as:

\be \label{norm_A}
\begin{split}
|\cA| ~&:=~ \sq{-g^{\m\n}\cA_\m \cA_\n} ~=~ \le( \fr{\f}{\fp} \ri)  \sq{-\hat{g}^{\m\n}\cA_\m \cA_\n}\\
& =~ 4\k \sq{3 q \cV_0} ~=~  4\k \sq{3 q ~\cU_0}
\end{split}
\ee
where $q = 1$ or $2$ depending on how the pseudo-trace mode was introduced in the system. For a detailed account and derivations of these results,  please refer \cite{ssasb2}.

\section{Einstein frame field equations} \label{sec:field_eqns}

 In terms of the scalar field $\vph$ and the matter lagrangian $\cL^{\le( m \ri)}$, the effective Einstein frame action is:

\be \label{action}
\cS_E ~=~ \int{d^4 x \sq{-g} \left[ \fr{\cR}{2\k^2} - \fr 1 {2} g^{\a\b} \pa_\a \vph ~ \pa_\b \vph - \cU(\vph) + \cL^{({\small m})}(g, \pa g, \vph) \right]}
\ee
where
\be \label{potential}
\cU(\vph) ~=~ \fr {2m^2}{\E^2}~ e^{-\E\vph}~~~~ , ~~~~ \E ~=~ 2\k\sq{\b}
\ee
Here $\b$ is a positive parameter used initially to couple the original Jordan frame scalar field  to the modified Ricci scalar. 
Also `$m$' is another positive constant. The action (\ref{action}) on first glance looks strikingly similar to that of a quintessence model in presence of matter. The difference lies in the fact that the matter lagrangian is non-minimally coupled to the scalar field in our case. This non-minimal coupling was picked up while conformally transforming the Jordan frame action. In doing so, although we successfully removed the non-minimal coupling between the scalar field and curvature, it was only at the expense of a similar coupling between the scalar field and the matter lagrangian. As a consequence of this, the matter and scalar field energy-momentum tensors are no longer individually conserved, a fact which is born out by the conservation equations of the system. The total energy-momentum tensor of the system is of course conserved. Using a Friedmann-Robertson-Walker (FRW) metric with the scale factor `$a$', the variation of action (\ref{action}) with respect to the metric yields:

\be \label{hubble}
H^2 ~\equiv~ \le(\fr{\dot{a}}{a}\ri)^2 ~=~ \fr{\k^2}{3} \le(\r_m + \r_{\vph} \ri)
\ee
and
\be \label{rchdri}
\dot{H} ~=~ - \fr {\k^2}{2} \le(\r_m + p_m + \r_\vph + p_{\vph} \ri)
\ee
A ($^{.}$) over a quantity represents it's time derivative. The conservation equations for the above system are as following \cite{frni}:

\be \label{conserv_matter}
\nab_\n T_{\m} ^{~\n(m)} ~=~ -\k \sq{\b} T^{(m)} \pa_\m \vph
\ee
and
\be \label{conserv_field}
\nab_\n T_{\m} ^{~\n(\vph)} ~=~ \k \sq{\b} T^{(m)} \pa_\m \vph
\ee
The quantities $T^{\m\n}$ and $T$ represent the energy-momentum tensor and it's trace for either of the two components, matter $(m)$ and scalar field $(\vph)$. The total energy-momentum tensor for the system is conserved, as is evident from addition 
of (\ref{conserv_matter}) and (\ref{conserv_field}). One can easily deduce from the action (\ref{action}) 
the energy density and pressure of the scalar field sector respectively as:

\be \label{energydensity_field}
\r_{\vph} ~=~ \fr 1 {2} {\dot{\vph}}^2 + \cU(\vph)
\ee
and
\be \label{pressure_field}
p_{\vph} ~=~ \fr 1 {2} {\dot{\vph}}^2 - \cU(\vph)
\ee
Also since matter is being modelled as dust, it has a vanishing pressure, i.e. $p_m=0$. With these inputs, the relations (\ref{conserv_matter}) and (\ref{conserv_field}) may be expanded as:



\begin{align}
\dot{\r}_{m } + 3H  \r_{m}    ~&=~ -\k ~\sq{\b} ~  \r_{ m} ~\dot{\vph} \label{conserv_matter2} \\
\dot{\r}_m + 3H \le( \r_{\vph} + p_{\vph} \ri) ~&=~ \k ~\sq{\b} ~  \r_{m} ~\dot{\vph} \label{conserv_field2}
\end{align}
Eqs. (\ref{hubble}), (\ref{rchdri}), (\ref{conserv_matter2}) and (\ref{conserv_field2}) are the field equations
 which determine the dynamics of our system. Only three of these are actually independent. For the scalar field potential of choice in the above model, this set of coupled differential 
 equations can infact be solved analytically. Such a solution and resulting dynamics of the universe have been discussed in 
 detail in \cite{ssasb2}. Here in this work we now take an alternative route to solving the field equations. An autonomous 
 system of first order coupled differential equations is constructed out of the available field equations. The system is 
 then analysed for the existence of real roots and their mathematical stability. 

\section{Phase plane analysis}  \label{sec:phase plane analysis}

\subsection{Constructing an autonomous system}

The first step in carrying out a phase plane analysis of our model is to construct an autonomous system of equations out of the existing field equations. It must be mentioned here that Eq. (\ref{conserv_matter2}) 
is infact a linear first order differential equation which can be solved explicitly to obtain matter energy density as 
a function of scale factor and the scalar field. Although straightforward to solve, we prefer to leave this equation out as the dependent equation and use Eqs. (\ref{hubble}), (\ref{rchdri}) and (\ref{conserv_field2}) as our working equations. By defining the phase-space variables:

\be \label{autonomous_coordinates}
X ~\equiv~ \fr{\k \dot{\vph}} {\sq{6} H} ~~~~~, ~~~~~ Y ~\equiv ~ \fr{\k \sq{\cU}}{\sq{3} H}
\ee 
the field equations can be reduced to the following autonomous equations:

\begin{align}
\fr{dX}{dN} ~&=~ -\fr 3 {2} X + \fr 3 {2} X \le(X^2 - Y^2 \ri) + \l Y^2 - \fr 3 {2} b \le(X^2 + Y^2 - 1 \ri) \label{autonomous_eqnx}  \\
\fr{dY}{dN} ~&=~ -\l XY + \fr 3 {2} Y \le(X^2 - Y^2 + 1 \ri) \label{autonomous_eqny}
\end{align}
and a constraint equation following directly from the Hubble's Eq. (\ref{hubble}):
\be \label{autonomous_eqnhubble}
X^2 + Y^2 + \fr {\k^2 \r^{(m)}}{3 H^2} ~=~ 1
\ee
Here $N\equiv ln(a)$ is the number of e-foldings. We have also defined:

\be \label{lambda,b}
\l ~=~ - \fr{\sq{6}}{2\k} ~ \fr 1 {\cU} ~\le( \fr{\pa \cU} {\pa \vph} \ri) ~~~~~, ~~~~~ b ~=~ \fr 2 {\sq{6}} ~ \sq{\b}
\ee
For the potential Eq.(\ref{potential}), it can be shown that:

\be \la{lambda_b_relation}
\l ~=~ 3b
\ee
The constraint Eq. (\ref{autonomous_eqnhubble}) can also be restated as:
\be \la{Omegarelation_field_m}
\O_{\vph} + \O_{(m)} ~=~ 1
\ee
with $\O_{\vph} = X^2 + Y^2$ and $\O_{(m)} = \le(1-X^2 - Y^2 \ri)$. The scalar field $\vph$ at any epoch can be found using the relation:
\be \la{vph}
\vph({N}) ~=~ \fr {\sq{6}}{\k}~ \cF \le( N \ri)   
\ee
where 
\be \la{integral_X}
\cF \le(N \ri) ~\equiv~ \int^{N} _0 X(\cN) d\cN
\ee
The explicit functional dependence of the variable $X$ on e-folds, needed to evaluate the above integral can in principle be found by solving Eqs. (\ref{autonomous_eqnx}) and (\ref{autonomous_eqny}).

But do the scalar field and matter sectors as obtained above constitute a good physical decomposition to study our universe? One may recall that by virtue of constraints (\ref{conserv_matter2}) and
(\ref{conserv_field2}) the scalar field and the matter sectors are mutually coupled. Neither sector can be treated exclusive of the other, hence rendering cosmological parameters like $\O_{(m)}$ and $\O_{\vph}$ of little to no physical significance when comparing with observed data. An effective decoupled system can be
obtained by singling out from the total energy density of the system  a part which diminishes by the third power of the scale factor ($a^{-3}$). This may be interpreted as the effective matter sector energy
density, $\rmf$ and the rest as the dark energy density ($\rX $)  \cite{ssasb2}. Building upon such a decomposition of the total energy budget of the universe, we obtain on integrating  Eq. (\ref{conserv_matter2}),
\be \label{rho_m_eff}
\rmf ~=~ \r _c - \rX ~=~ \r_m ~e^{\k \sq{\b} \vph}
\ee
The effective matter density parameter is given as:
\be \label{Omega_m_eff}
\begin{split}
\Omf ~&\equiv~ \fr {\k^2 \rmf} {3 H^2} \\
&= \le(1- X^2 - Y^2 \ri) e^{3b \cF}
\end{split}
\ee
Eq. (\ref{rho_m_eff}) dictates that,
\be
\OX ~=~ 1 - \Omf
\ee
The equation of state (E.o.S.) parameter for the dark energy sector,
\be \label{omega_de}
\begin{split}
\oX ~&\equiv~ \fr {p_{\!_{X}}} {\r_{\!_{X}}}\\
&= \fr {X^2 - Y^2} {\OX}
\end{split}
\ee
and finally the total E.o.S. parameter for the system,
\be \label{omega}
\begin{split}
\o ~&\equiv~ \oX ~\OX \\
&=  X^2 - Y^2 
\end{split}
\ee
Before we proceed to find solutions for the above system of equations, there are some deductions that can be made from inspecting these equations:

\bi

\item The system constituted by Eqs. (\ref{autonomous_eqnx})- (\ref{autonomous_eqnhubble}) is symmetric under the transformation $Y \longrightarrow -Y$. As such we can restrict our analysis to the region of phase plane where $Y \geq 0$ without 
any loss of generality.

\item The set of equations defining the above autonomous system are almost same as those characterising a 
quintessence field in presence of dust. The difference appears 
as the last term in the right hand side of Eq. (\ref{autonomous_eqnx}). In-fact the origin of this term can be traced 
to the coupling term in Eq. (\ref{conserv_field2}). It can be shown that:
\be
\fr 3 {2} b \le(1 - X^2 - Y^2  \ri) ~=~ \fr 1 {\sq{6} H^2} ~ \k^2 \sq{\b} \r^{(m)}
\ee
Omitting this term from the analysis (equivalently setting $b=0$) is equivalent to the removal of the non-minimal coupling between the scalar field and the 
matter lagrangian and consequently any interactions between the two sectors.
\ei

Also one can co-relate the dynamics of the system with  the norms of the individual torsion modes ($|\cT|^2$ and $|\cA|^2$) as: 

\be \la{torsion_norms_einstein}
\fr{|\cT|^2}{H^2} ~=~ 27b^2 e^{6b \cF} X^2 ~~~,~~~ \fr{|\cA|^2}{H^2}  ~=~ 144 q ~e^{6b \cF}  Y^2
\ee
here $q = 1$ or $2$ depending upon the the two schemes used in \cite{ssasb2} to deal with the pseudo-trace mode of torsion.

Analysing the above autonomous system for stable solutions (the theory behind this analysis is reviewed in the Appendix), we find there exist seven critical points (five if accounted for multiplicity) for our autonomous system. These points 
along with their domains of existence and physical relevance are listed in Table \ref{table_roots}. Also listed are some cosmological parameters at each of these critical points.
The eigenvalues of the matrix $\cM$ at each critical point, the types of critical points and their nature are tabulated in Table \ref{table_eigenvalues}. 
\begin{table}[h] 
\centering
\renewcommand{\arraystretch}{1.5}
\begin{tabular}{||c|c|c|c|c|c|c|c||}
\hline
\multirow{2}{*}{Name} & \multirow{2}{*}{$X_c$} & \multirow{2}{*}{$Y_c$}& Domain of  & Domain of physical  &  \multirow{2}{*}{$\O_{DE}$} & \multirow{2}{*}{$\o_{\!_{DE}}$} & \multirow{2}{*}{$\o$} \\ 
  &   &   & existence & relevance &  &  &  \\
\hline\hline
$\#~1$ & -1 & 0 & $ b \in (0, \infty)$  & $ b \in (0, \infty)$  &  $1$ & $1$ & $1$ \\ 
\hline
$\#~2$ & 1 & 0 & $ b \in (0, \infty)$ &  $ b \in (0, \infty)$ &  $1$ & $1$ & $1$ \\
\hline
$\#~3$ & $b$ & 0 & $b \in (0, \infty)$  & $ b = 1 $ &  $b^2$ & $1$ & $b^2$ \\
\hline
$\#~4$ & $b$ & $ \pm \sq{1-b^2} $ & $ b \in (0, 1]$ & $ b \in (0, 1]$ &   $1$ & $2 b^2 - 1$ & $2 b^2 - 1$ \\
\hline
$\#~5$ & $\fr 1 {b}$ & $ \pm \sq{\fr 1 {b^2} - 1} $ & $b \in (0, 1]$  & $ b = 1 $  &  $1$ & $ 1$ & $ 1$\\
\hline
\end{tabular}
\renewcommand{\arraystretch}{1}
\caption{ 
{\footnotesize The critical points for the autonomous system under consideration along with the domain of the parameter $b$ for which these points exist and yield a physically compatible value of effective matter energy density. Also listed are cosmological parameters at each of these critical points.} }
\label{table_roots}
\end{table}

\begin{table}[h]
\centering
\renewcommand{\arraystretch}{1.6}
\begin{tabular}{||c|c|c|c|c||}
\hline
Name & $\m_1$ & $\m_2$ & Type & Nature \\
\hline\hline
$\#~1$ & $3+3b $ & $3+3b $ & Nodal Source $ \forall ~ b \in (0, \infty)$ & Unstable  \\
\hline
\multirow{2}{*}{$\#~2$} & \multirow{2}{*}{$3-3b $} & \multirow{2}{*}{$3-3b $} & Nodal Source for $ \forall ~ b \in (0, 1)$ & Unstable  \\
   &  &   & Nodal Sink for $ \forall ~ b \in (1, \infty)$ &  Stable   \\
\hline
\multirow{2}{*}{$\#~3$} & \multirow{2}{*}{$\fr 3 {2} \le(1-b^2 \ri) $} & \multirow{2}{*}{$- \fr 3 {2} \le(1-b^2 \ri) $} & Saddle point for $ \forall ~ b \in (0, 1)$ & Unstable  \\
   &  &   & Indeterministic for $ b = 1$ &  Indeterministic   \\
\hline
$\#~4$ & $ -3+3b^2  $ & $ -3+3b^2  $ & Nodal Sink for $ \forall ~ b \in (0, 1)$  & Stable   \\
\hline
$\#~5$ & $ 3b\le( 1 - \fr 1 {b^2}\ri)  $ & $ -3b\le( 1 - \fr 1 {b^2}\ri)   $ & Indeterministic   &  Indeterministic   \\
\hline
\end{tabular}
\renewcommand{\arraystretch}{1}
\caption{{\footnotesize The eigenvalues of matrix $\cM$ corresponding to each critical point along with the type and nature of the critical point.} }
\label{table_eigenvalues}
\end{table}

\subsection{Dynamics of the universe at each critical point}

For our system to be physically consistent with observations, the effective matter density parameter must satisfy the constraint 
\be \la{Omega_m_inequality}
0 \leq \Omf \leq 1
\ee
at all times. But it is evident from Eq.(\ref{Omega_m_eff}) that the the density parameter keeps evolving even after the system has reached a critical point due the contribution coming from the ever evolving exponential factor. The constraint (\ref{Omega_m_inequality}) is hence violated sooner or later rendering the model un-physical at that critical point. The only exception occurs when the critical point obeys the relation:
 
\be \la{critpt_existence}
X_c ^2 + Y_c ^2 = 1
\ee
effectively making $\Omf =0$ in the asymptotic limit. Hence only those critical points in the phase space which lie on the circumference of the circle centred at the origin having a unit radius are relevant for our work. This must be emphasised as a key difference between our model and those where the scalar field and matter remain decoupled. In our case only those cosmological solutions where the matter sector completely dies out with time are permissible.  

Also the parameter `$b$' can only 
take positive values. The case $b=0$ is also not permitted by our original model. Infact a  vanishing `$b$' would 
kill the interaction term in (\ref{conserv_matter}) and (\ref{conserv_field}) yielding the field equations for the standard quintessence model which has already been extensively studied in literature. With these points in mind, we now proceed to analyse 
the physical aspects of the universe and it's constituents at different critical points.

\bi
\item Critical point \# 1: This critical point exists for all values of the parameter $b$. 
The corresponding eigenvalues are always real and positive, making this point an unstable nodal sink. 
The solutions at this point are dark energy dominated ($\OX = 1$), which exhibits a steep fluid like behaviour  ($\oX = 1$). Since $\o > - 1/3$ at this point, it can be inferred that the universe is in a state of  de-acceleration. 

\item Critical point \# 2: This critical point again exists for all values of $b$, behaving as an unstable 
nodal source whenever $b<1$ and as a stable nodal sink when $b >1$. The solutions in either case are 
dominated by a steep fluid like dark energy component leading to a de-accelerating universe. 

\item Critical point \# 3: Although this critical point exists whenever $b \leq 1$, in view of Eq. (\ref{critpt_existence}) it is physically relevant only when $b=1$. Here it coincides with the critical point \# 2. This point behaves as a saddle point whenever $b \neq 1$. It's behaviour when $b=1$ remains indeterministic as both the eigenvalues vanish simultaneously.

\item Critical point \# 4: This critical point moves on the circumference of the circle $ X^2 + Y^2 = 1$ and exists 
whenever $b \leq 1$, coinciding with critical point $\#2$ when $b=1$. Behaving as an asymptotically stable nodal sink 
whenever $b<1$, it leads to a dark energy dominated universe which can be in a state of acceleration or de-acceleration 
depending on whether $b < (1/\sq{3})$ or otherwise respectively. This is a unique critical point in our system as it
is the only one which can support solution (stable or otherwise) in presence of finite scalar field potential energy.

\item Critical point \# 5: This critical point exists whenever $b\leq 1$ but is relevant only when $b=1$, which makes it coincident with the 
critical point $\#2$.  
\ei
The critical points \#~3 and \#~4 showcase another key difference between our model and that of a quintessence scalar field in presence of dust. In the latter model, the corresponding critical points lie permanently at points ($0,0$) and ($0,1$) respectively (taking the limit $b \rightarrow 0$ of
our model). The saddle point there leads to solutions where the scalar field eventually dies out leaving matter sector as the sole constituent of the universe. The stable point there represents the $\L$CDM solutions which are completely dominated by a non-dynamic dark energy sector. The critical point \#~3 in
our case does permit co-existence of the matter and dark energy sectors. The solutions at critical point \#~4 on the other hand are again dominated by the dark energy. But the dark energy sector here has a certain dynamicity associated with it.

We have thus established the existence of stable points as equilibrium solutions to the set of autonomous equations 
representing our original MST model in the phase plane. But is the dynamics of the universe and it's 
constituent components, dictated by these stable points in concordance with solutions obtained by explicitly solving the field equations analytically in our original work \cite{ssasb2}? Clearly the universe as pictured by our analytic solutions 
lies hidden in  dynamics of the dark energy and matter at one of the five critical points. It is just a question of identifying the right one. Right from the beginning critical point $\# 4$ seems the prime (or rather the sole) contender as this is 
the only point whose solutions can accommodate an accelerating universe that we seek. Using the definition of the phase plane coordinates, Eq. (\ref{autonomous_coordinates}), at this critical point,

\be
\fr{\k \dot{\vph}} {\sq{6} H}~=~b  ~~~~~and ~~~~~  \fr{\k \sq{\cU}}{\sq{3} H} ~=~ \sq{1-b^2}
\ee
Some trivial mathematics along with Eq.(\ref{potential}) reveals the evolution of the scalar field  and the Hubble parameter  
in terms of the scale factor, in a universe whose dynamics is dictated by the critical point $\# 4$ to be,

\be
H ~=~ \fr {m^2}{9 ~ b^2 \le(1 - b^2 \ri)} ~ a^{-3 b^2}
\ee
and
\be
e^{- \E \vph}  ~=~ a^{-6 b^2}
\ee
In view of Eq.(\ref{lambda,b}) these results are exactly in line with our analytical solutions. Let us also look qualitatively into the type of universe predicted by the critical point $\# 4$. To this end, it is better to fall back on to the coupled system of the scalar field ($\vph$) and matter as it provides a more intuitive picture of the dynamics taking place in the phase space. Any point on the $Y$ axis 
of the phase space relates to a scalar field configuration which has only a potential term and no kinetic term. As such the scalar field mimics the 
cosmological constant having it's E.o.S. parameter as $-1$. What this translates to in terms of individual torsion modes is the case when only the pseudo scalar mode of torsion exists without the trace mode (which has an implicit dynamicity in terms of the scalar field). The energy density derived from the pseudo scalar mode does the job of the cosmological constant.  Also interpreting Eq.(\ref{autonomous_eqnhubble}) geometrically, the further a point is from the origin (of course being subjected to constraint $X^2 + Y^2 \leq 1$ at all times), the greater 
is the scalar field (and hence torsion) contribution to the overall energy content of the universe. Hence the standard $\L$CDM trajectory 
moves along the $Y$ axis and approaches asymptotically the point $(0,1)$ situated at the boundary of the permissible 
region of the phase plane. This solution can indeed be recovered using our model in the limit $b \rightarrow 0$.

Now any point on the phase plane with a non-zero value of $X$ represents a system configuration where scalar field has an associated dynamicity, strength of which is again determined by magnitude of $X$ itself. 
Hence for $b <<1$, the critical point $\# 4$, which again lies on the boundary, describes a cosmology which can be summed up as small fluctuations over the $\L$CDM type cosmology. The fluctuations arise due to the small dynamicty that the scalar field 
sector now enjoys. This can also be seen as the torsion trace mode now coming into existence, having it's energy density alter the $\L$CDM cosmology 
on part of the pseudo trace mode. Such a cosmology concurs with our analytical results. A small value for $\b$ and consequently for $b$ was infact inferred in our original work using data from cosmological observations. 

The quantitative behaviour of $H$ and $\vph$ and the qualitative nature of the predicted cosmology allow us to say with certainty that the critical point $\# 4$ represents 
our analytical solutions in the Einstein frame and that these solutions are stable in nature, subject to restrictions on the model parameter. 
It now remains to constrain the model parameter `$b$' using the results we got. There are two basic inequalities that must 
be satisfied at all times for the critical point$\# 4$ to represent a cosmology consistent with observations. These are:

\ben
\item Existence condition: One must ensure that the critical point in question exists in the first place.  Table \ref{table_roots} lists the corresponding condition as,
\be
b ~\leq ~1
\ee

\item Accelerating universe condition: As is well known, the universe is in a state of acceleration. This means that the total effective E.o.S. parameter must satisfy the condition, $\o < -1/3$. Using results given in Table \ref{table_roots} this gives,
\be
\le(2 b^2  - 1\ri) ~<~ -\fr 1 {3}
\ee
implying
\be
b~ < \fr 1 {\sq{3}}
\ee
\een
The latter of the two constraints provides a tighter bound on the possible set of values the free parameter can take. In view of Eq. (\ref{lambda,b}), this constraint can be stated as:
\be \label{constraint_beta_einstein}
\b < \fr 1 {2}
\ee

In the concluding section, we subject our autonomous equations to numerical recipes in order to cross check 
the equilibrium solutions we obtained and also to extract the exact trajectories a system takes in the phase plane to reach those solutions. 
We also note the evolution of some cosmological parameters of interest over the lifetime of a trajectory.

\section{Solutions using numerical methods} \label{sec:numeric}

\begin{figure}[h]
\centering
   \begin{subfigure}{0.49\linewidth} \centering
     \includegraphics[scale=0.28]{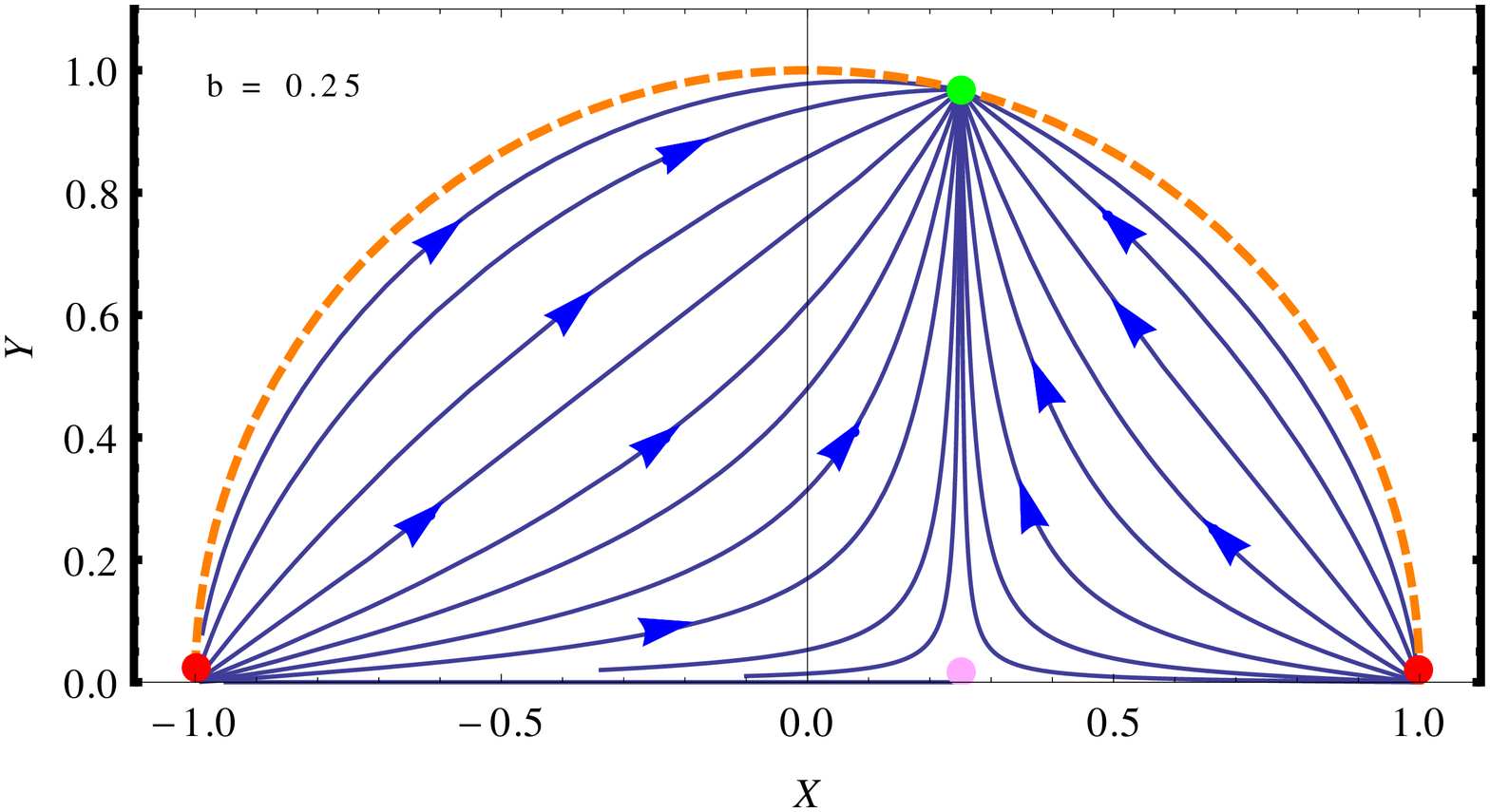}
     \caption{{\footnotesize The ($X-Y$) plane when $b=0.25$}}\label{fig:fig1A}
   \end{subfigure}
   \begin{subfigure}{0.49\linewidth} \centering
     \includegraphics[scale=0.28]{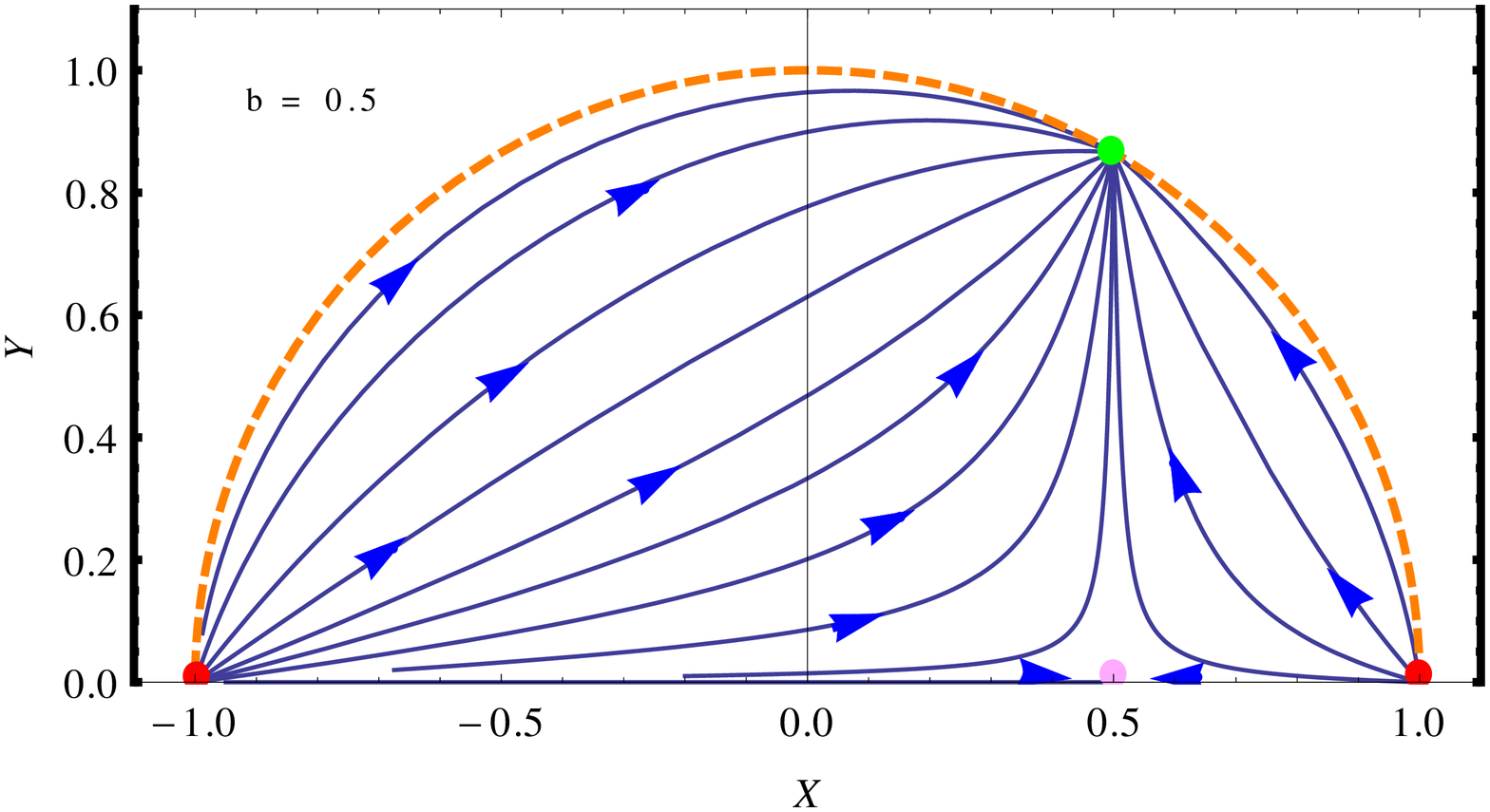}
     \caption{{\footnotesize The ($X-Y$) plane when $b=0.50$}}\label{fig:fig1B}
   \end{subfigure}
   \begin{subfigure}{0.49\linewidth} \centering
     \includegraphics[scale=0.28]{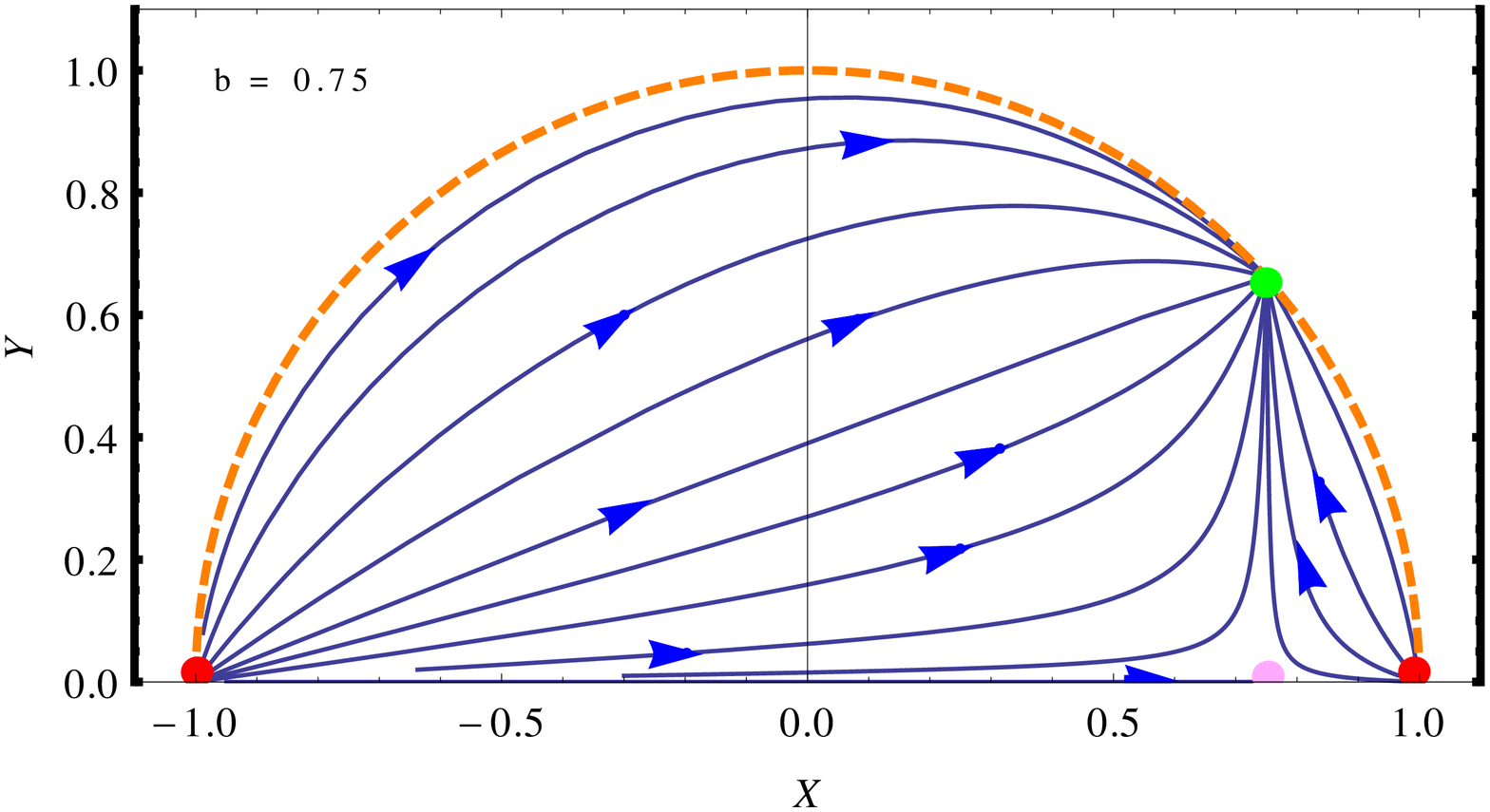}
     \caption{{\footnotesize The ($X-Y$) plane when $b=0.75$}}\label{fig:fig1C}
   \end{subfigure}
   \begin{subfigure}{0.49\linewidth} \centering
     \includegraphics[scale=0.28]{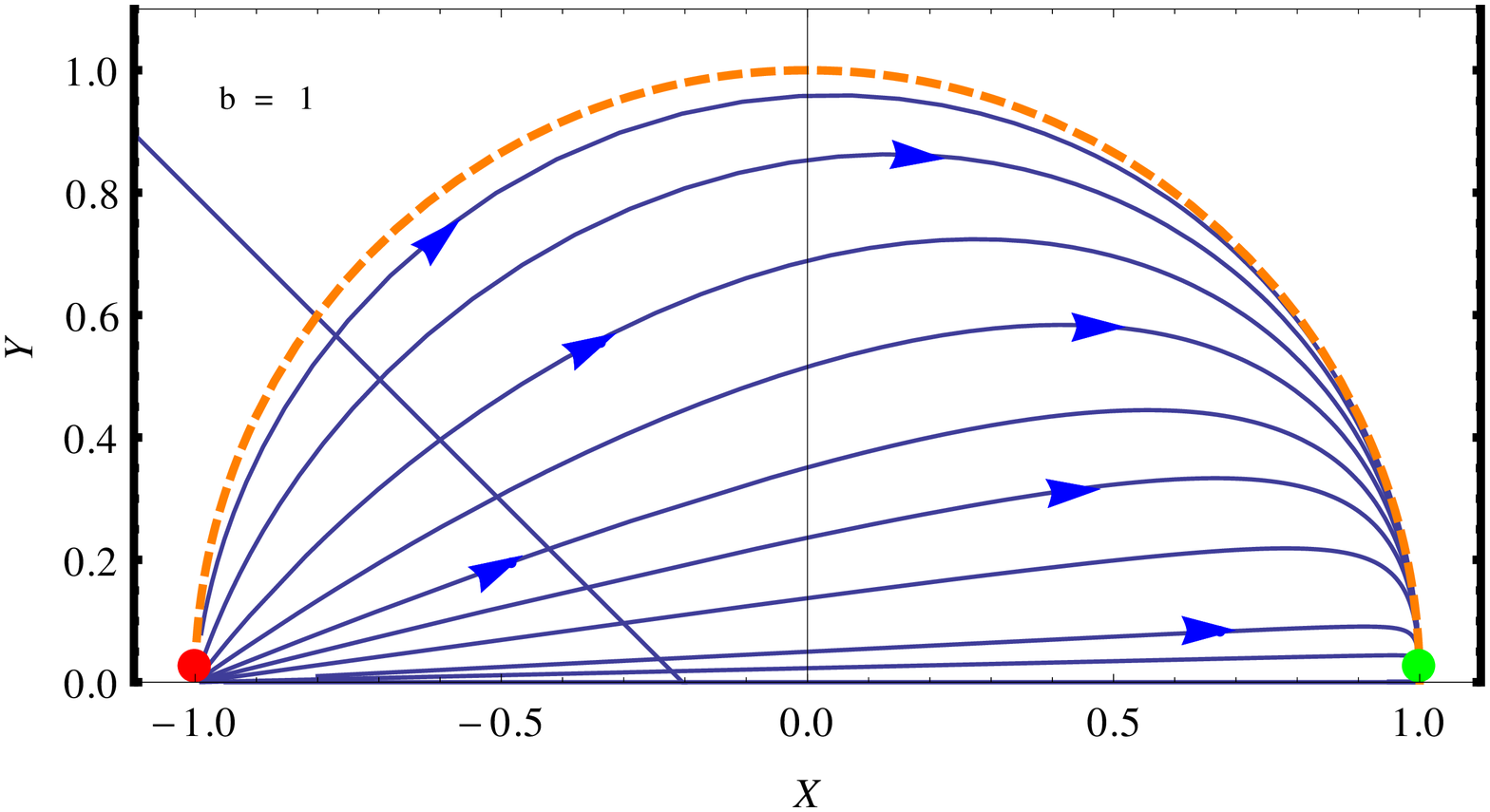}
     \caption{{\footnotesize The ($X-Y$) plane when $b=1.00$}}\label{fig:fig1D}
   \end{subfigure}
\caption{{\footnotesize The phase portraits for four different values of the parameter $b$. The dots represent the critical points, arrows mark the direction of evolution of trajectories 
over time and the dashed curve demarkates the region of phase plane which supports cosmologies having a non-negative matter energy density. As we increase the value of $b$, the saddle and the stable points of the system shift right along the abscissa and the circumference of the unit circle respectively till they coincide with a third ciritical point at $X=1$ when $b=1$.  }} \label{fig:fig1}
\end{figure}

The analysis and results presented so far have been based on the equilibrium solutions of the autonomous set of equations representing our model in a phase plane. These equilibrium solutions describe the system only at a particular critical point. They fail to give any information, atleast
quantitatively, about the state a system is in before it attains the critical solutions. To overcome this limitation (atleast partially), we turn to numerically solving the set of autonomous Eqs. (\ref{autonomous_eqnx})-(\ref{autonomous_eqnhubble}). Although numerical methods fail to return
explicit solutions for $X(N)$ and $Y(N)$ in a closed form, given a set of initial values of the phase space variables $X$ and $Y$, they do manage to
trace the trajectory of the system as it evolves in the phase plane, right from the time it originates at the aforementioned coordinates till the point when it terminates at one of the critical points. 
These methods also allow us to plot the evolution of any function depending explicitly on the phase plane coordinates, over the lifetime of a
trajectory. Hence the evolution of quantities such as $\OX$ , $\o$ and $\oX$ in a cosmology resulting from the chosen initial conditions can be plotted practically over the whole time span of the universe. 
In Fig. \ref{fig:fig1A} - Fig. \ref{fig:fig1D} we plot the trajectories our system takes in the ($X-Y$) plane to reach a critical point for
different initial values when $b=0.25$, $0.50$, $0.75$ and $1.00$ respectively.   Superimposed for comparison 
are the critical points we found in the last section. 

The trajectory that the system takes in the phase plane must at no point over it's lifetime cross the circumference of the unit circle centred at the origin as this would render the matter density parameter (\ref{Omega_m_eff}) negative. It is seen that trajectories originating anywhere 
on the valid phase plane except the abscissa terminate at the critical point $\# 4$. The trajectories originating at the abscissa end their journey at the 
critical point \# 3, which as discussed earlier acts as a saddle point when $b<1$. This makes $Y=0$ the stable axis for the saddle point and $X=b$ the unstable axis. Hence trajectories originating anywhere except the $X$ axis are 
deflected towards the critical point $\# 3$ in the horizontal direction whereas in the vertical direction they are deflected 
away from it. The saddle point hence has the effect of funnelling trajectories towards the stable critical point $\# 4$, which is always situated vertically above itself in the phase plane. Of course the sink
exercises it's own attractive power as well. Although the positions of the stable and saddle points in the phase plane depend on the model parameter, $b$ ($X_c = b$) for both these points throughout their domain of existence), the above reasoning retains it's validity as the two critical points are always situated
vertically one over the other. As is evident, increasing the value of the model parameter shifts the saddle and stable points to their right along the abscissa and the circumference of the unit circle
respectively. The shift continues till $b=1$ when these critical points coincide with a third critical point situated at $X=1$. Further increase in the value of $b$ takes the critical point \#~3 outside the physical realm while the critical point \#~4 ceases to exist from here on.

\begin{figure}[h]
\begin{minipage}[t]{0.48\textwidth}
\includegraphics[width=\linewidth,keepaspectratio=true]{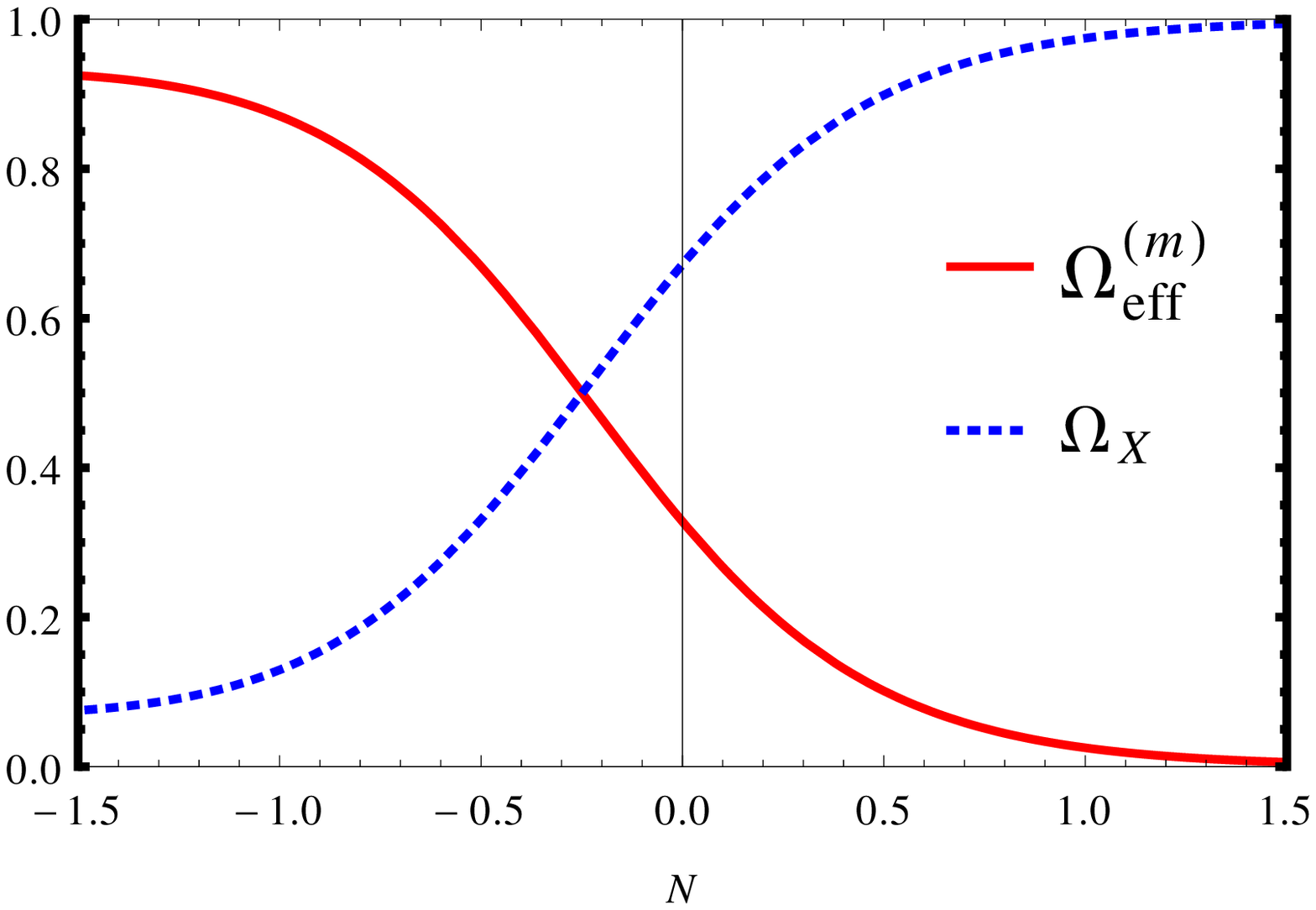}
\caption{{\small Evolution of the dark energy density parameter $\OX$ and the effective matter density parameter $\Omf$ over the trajectory with initial conditions $\le( X_i \equiv 0, Y_i \equiv 5 \times 10^{-7}\ri)$ and $b=0.1$.} }
\label{fig2}
\end{minipage}
\hspace*{\fill} 
\begin{minipage}[t]{0.48\textwidth}
\includegraphics[width=\linewidth,keepaspectratio=true]{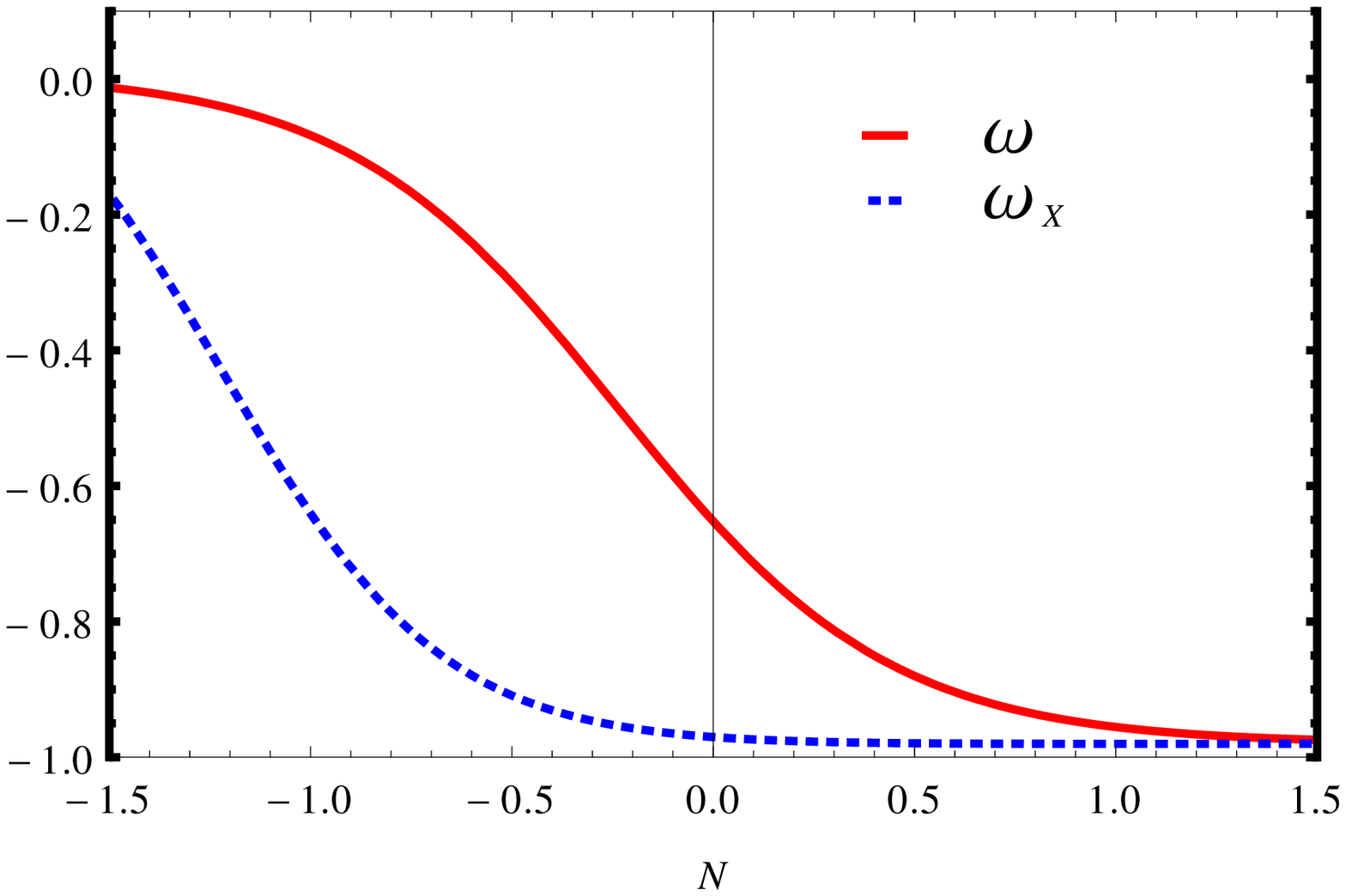}
\caption{{\small Evolution of the total EoS parameter for the universe, $\o$ and the EoS parameter for the dark energy sector, $\oX$ over the trajectory with initial conditions $\le( X_i \equiv 0, Y_i \equiv 5 \times 10^{-7}\ri)$ and $b=0.1$.} }
\label{fig3}
\end{minipage}
\end{figure}

\begin{figure}[h]
\begin{minipage}[t]{0.48\textwidth}
\includegraphics[width=\linewidth,keepaspectratio=true]{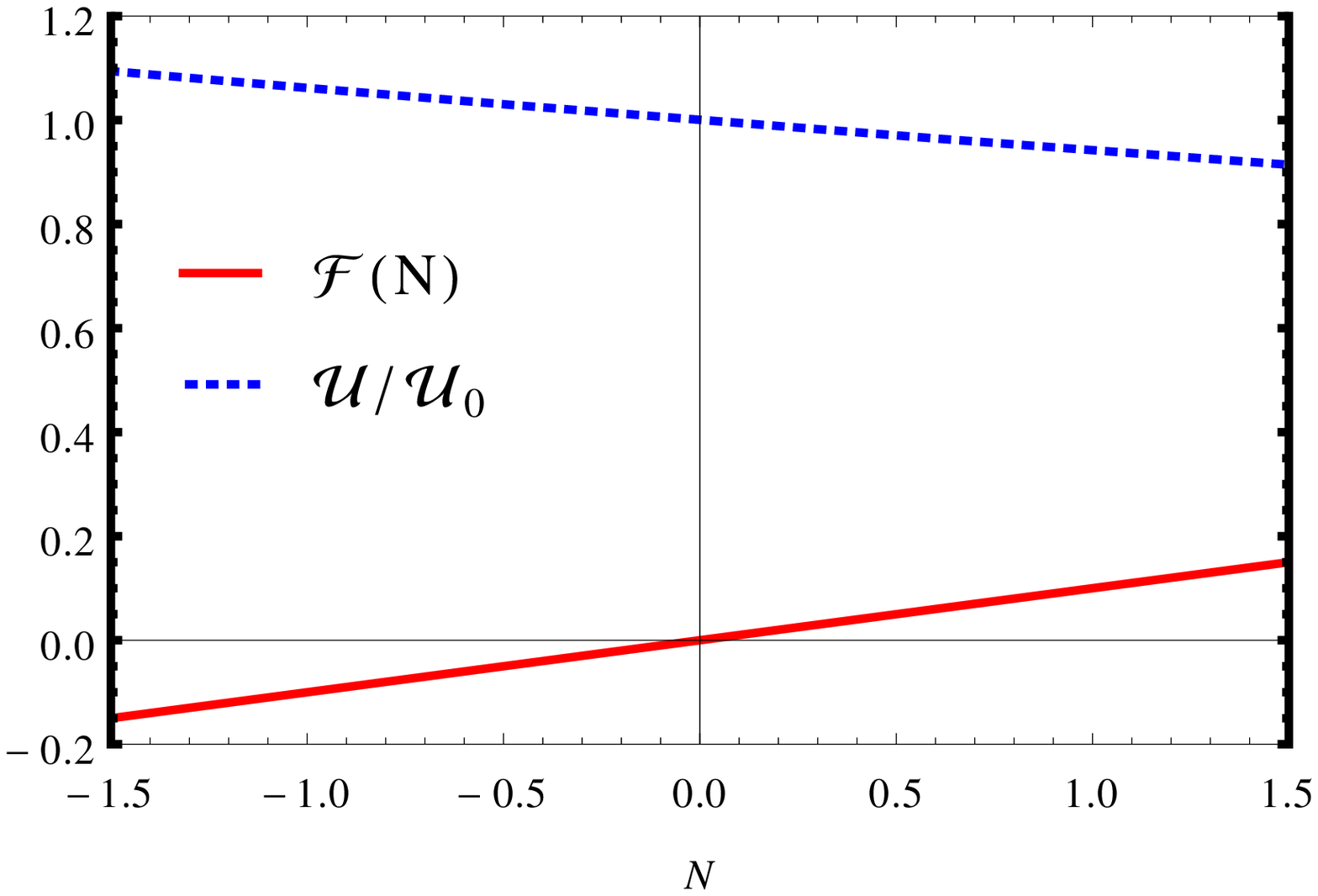}
\caption{{\small Evolution of $\cF(N)$ and the normalized scalar field potential  $\cU / \cU_0$ over the trajectory with initial conditions $\le( X_i \equiv 0, Y_i \equiv 5 \times 10^{-7}\ri)$ and $b=0.1$.} }
\label{fig4}
\end{minipage}
\hspace*{\fill} 
\begin{minipage}[t]{0.48\textwidth}
\includegraphics[width=\linewidth,keepaspectratio=true]{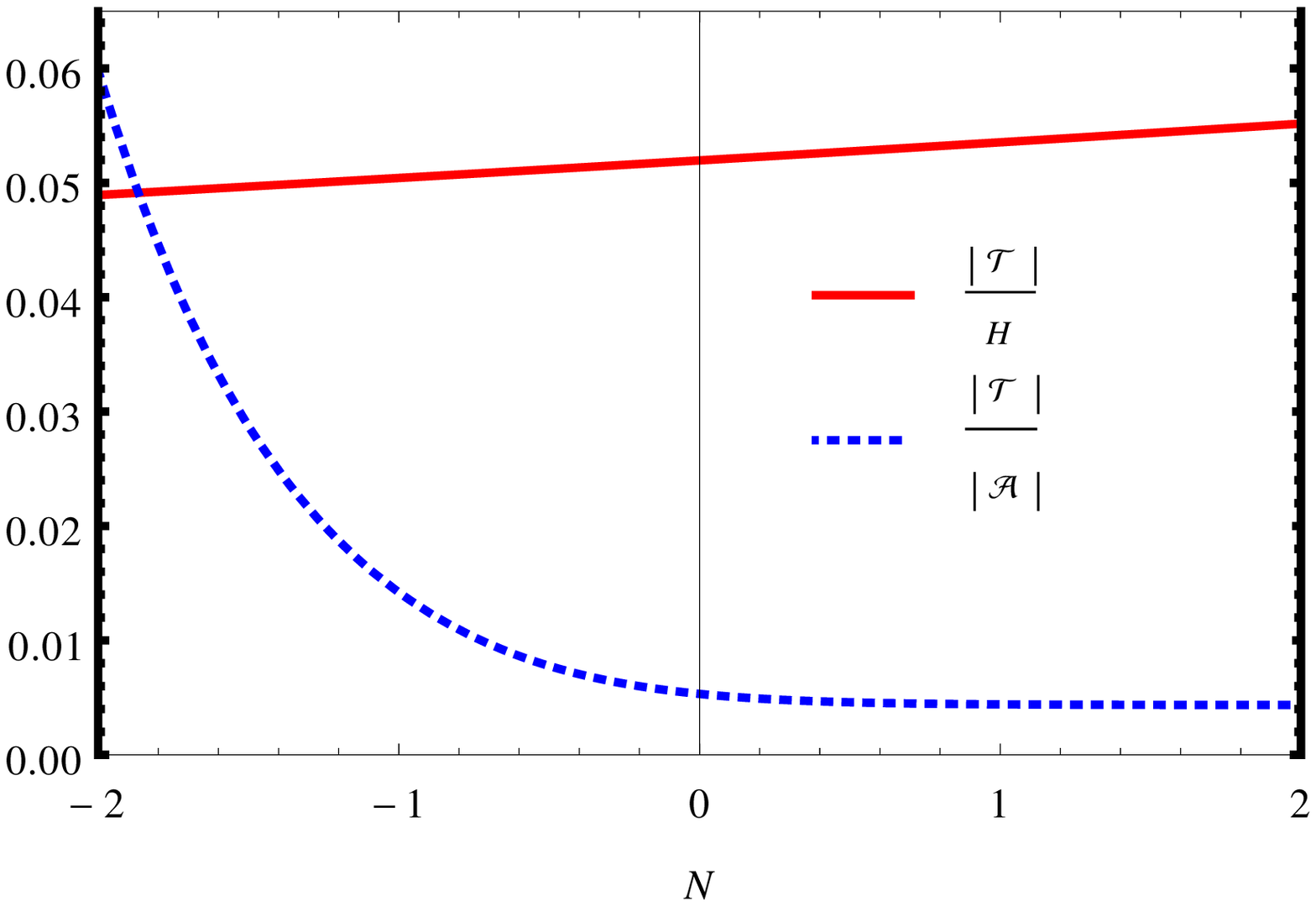}
\caption{{\small Norm evolution of the trace mode of torsion, $|\cT|$ in units of $H$ and the relative strengths of the norms of the two torsion modes over a trajectory with initial conditions $\le( X_i \equiv 0, Y_i \equiv 5 \times 10^{-7}\ri)$, $b=0.1$ and $q=1$.} }
\label{fig5}
\end{minipage}
\end{figure}

Plotted in Fig. \ref{fig2} are the evolutions of $\Omf$ and $\OX $ over one fiducial trajectory with initial
conditions, $\le( X_i \equiv 0, Y_i \equiv 5 \times 10^{-7} \ri)$ when $b=0.1$. Initial conditions are so chosen to allow for the dominance of the effective matter sector during the early epochs. It can be seen that the evolution of the density parameter of either sector, over the particular
trajectory in question is very nearly same as that predicted by the $\L$CDM model. The key difference lies in the nature of the dark energy.  Our model allows for a dynamic dark energy sector, which is evident from the evolution of dark energy EoS parameter, $\oX$ in Fig.  \ref{fig3} plotted alongside the total EoS parameter of the system, $\o$, for the same set of initial conditions. Fig. \ref{fig4} shows the evolution
over time of the normalized scalar field potential, ${\cU}/{\cU_0}$ and $\cF(N)$ as defined by Eq. (\ref{integral_X}). It is evident that there is no appreciable change in the value of the potential in the interval of time being considered. However slow in it's evolution, it must be remembered that it is this potential energy that drives the system towards the stable critical point. 

The evolution over time of the torsion trace norm (in units of $ H$) and the relative strengths of the two torsion norms ($|\cT| / |\cA|$) are shown in Fig. \ref{fig5}. This ratio diminishes sharply with time, saturating asymptotically to a point where the strength of the pseudo-trace mode is about two order of magnitudes stronger than that of the trace mode. This is hardly surprising as it can be shown using (\ref{torsion_norms_einstein}) that the relative strengths of the two modes asymptotically go as:
\be
\fr {|\cT|}{|\cA|} \sim \fr {b^2} {\sq{1-b^2}}
\ee
A small value of the parameter $b$, which is essential for maintaining consistency with cosmological observations, hence keeps this ratio small. In the next section we deal with the Jordan frame formulation of our model and investigate the mathematical stability of the analytical solutions in that frame.

\section{Phase plane analysis in the Jordan frame} \label{sec:jordan}

The MST model was conceived in the Jordan/Conformal frame with a characteristic non-minimal coupling between the scalar field and the Ricci scalar. We presented the Einstein frame analysis first as it was straightforward and exactly in line with the canonical formulation of Einstein's General Relativity. Having established the stability of our Einstein frame solutions, we now turn our attention to their Jordan frame  counterparts. In a very concise fashion, we present the mathematical formulations and some important results for this frame. 

As mentioned already, in Jordan frame our model reduces to the Brans-Dicke (BD) model with a BD parameter, $\fw$.  The effective action for the model in Jordan frame may be written as:
 
\be \label{action_jordan}
\cS_J ~=~ \int{d^4 x \sq{-g}\le[\fr{\F}{2} \cR ~- \fr {\fw}{2\F} g^{\m\n} \pa_{\m} \F ~\pa_{\n} \F ~- \cV \le(\F \ri) + \cL^{({\small m})}(g, \pa g)     \ri]}
\ee
where $\F$ is a redefined scalar field in this frame. The following relations hold true for the model under consideration:
\be \label{BDparameter}
\fw~=~ \le(\fr 1 {4\b} - \fr 3 2 \ri) ~~~~~,~~~~~~  \cV \le(\F \ri) ~=~ \fr{m^2}{2\b} ~ \F
\ee
Since the parameter $\b $ can take only positive non-zero values, it can be concluded from the above relation that:
\be
-\fr 3 2 ~< ~\fw ~ <~ \infty
\ee
Representing our space-time by the FRW metric, the field equations for the system can be summarized as:

\begin{align}
3 \F H^2 ~=~ \r_m +~ \fr{\fw}{2\F} ~\le(\dot{\F}\ri)^2  + \cV - 3H\dot{\F} \label{JordanEom1} \\
2\F \dot{H} ~=~ -\r_m ~- \ddot{\F}~ - \fr{\fw}{\F} ~\le(\dot{\F}\ri)^2 + H \dot{\F} \label{JordanEom2}\\
\cR ~=~ 2\fw \le( \fr {\ddot{\F}}{\F} ~+~ 3H \fr{\dot{\F}}{\F} ~-~  \fr{\le(\dot{\F}\ri)^2}{2\F^2} \ri) + 2~ \fr{\pa \cV}{\pa \F} \label{JordanEom3}
\end{align}
and
\be
\dot{\r}_{\le(m \ri)} + 3H  \r_{(m)}    ~=~ 0 \label{JordanEom4}
\ee
Only three out of the above four field equations are actually independent. Defining the phase space variables \cite{tsujikawa},

\be
X ~\equiv ~ \fr {\dot{\F}} {\sq{6} H \F} ~~~~~,~~~~~ Y ~\equiv~ \fr 1 H \sq{\fr{\cV}{3\F}}
\ee
the field Eqs. (\ref{JordanEom1})-(\ref{JordanEom3}) reduce to:

\begin{align}
\fr{dX}{dN} ~&=~ -\fr {\sq{6}} {2\le (2 \fw +3 \ri)} \le[\le(5 \fw +6\ri) X^2  - \le(3-2 \l \ri) Y^2 + 2\sq{6} \le(1+ \fw \ri) X - 1\ri] - \fr{\dot{H}}{H^2} X \label{Jordan_autonomous_eqnx}\\
\fr{dY}{dN} ~&=~  \fr {\sq{6}} {2} \le(\l - 1 \ri) XY ~- \fr{\dot{H}}{H^2} Y \label{Jordan_autonomous_eqny}
\end{align}
with
\be
\fr {\dot{H}}{H^2} ~=~ \fr 3 {\le(2 \fw +3 \ri)} \le[ -\fw \le(1+\fw \ri)X^2 + \fr {\sq{6}}{3} \fw X + \le(\l + \fw \ri)Y^2 - \le(2 + \fw \ri) \ri]
\ee
The parameter $\l$ in this frame is defined as:
\be
\l ~\equiv~ \fr {\F}{\cV}  \le( \fr{\pa \cV}{\pa \F} \ri)
\ee
In this frame, the phase space variable and norms of individual torsion modes (in units of Hubble parameter) are related as:

\be
\fr {|\cT|^2}{H^2} ~=~ \fr {27}{2} X^2 ~~~,~~~ \fr {|\cA|^2}{H^2} ~=~ 144 ~q  ~Y^2
\ee
where $q$ as before can take values $1$ or $2$.

\begin{figure}[h]
\centering
   \begin{subfigure}{0.45\linewidth} \centering
     \includegraphics[scale=0.35]{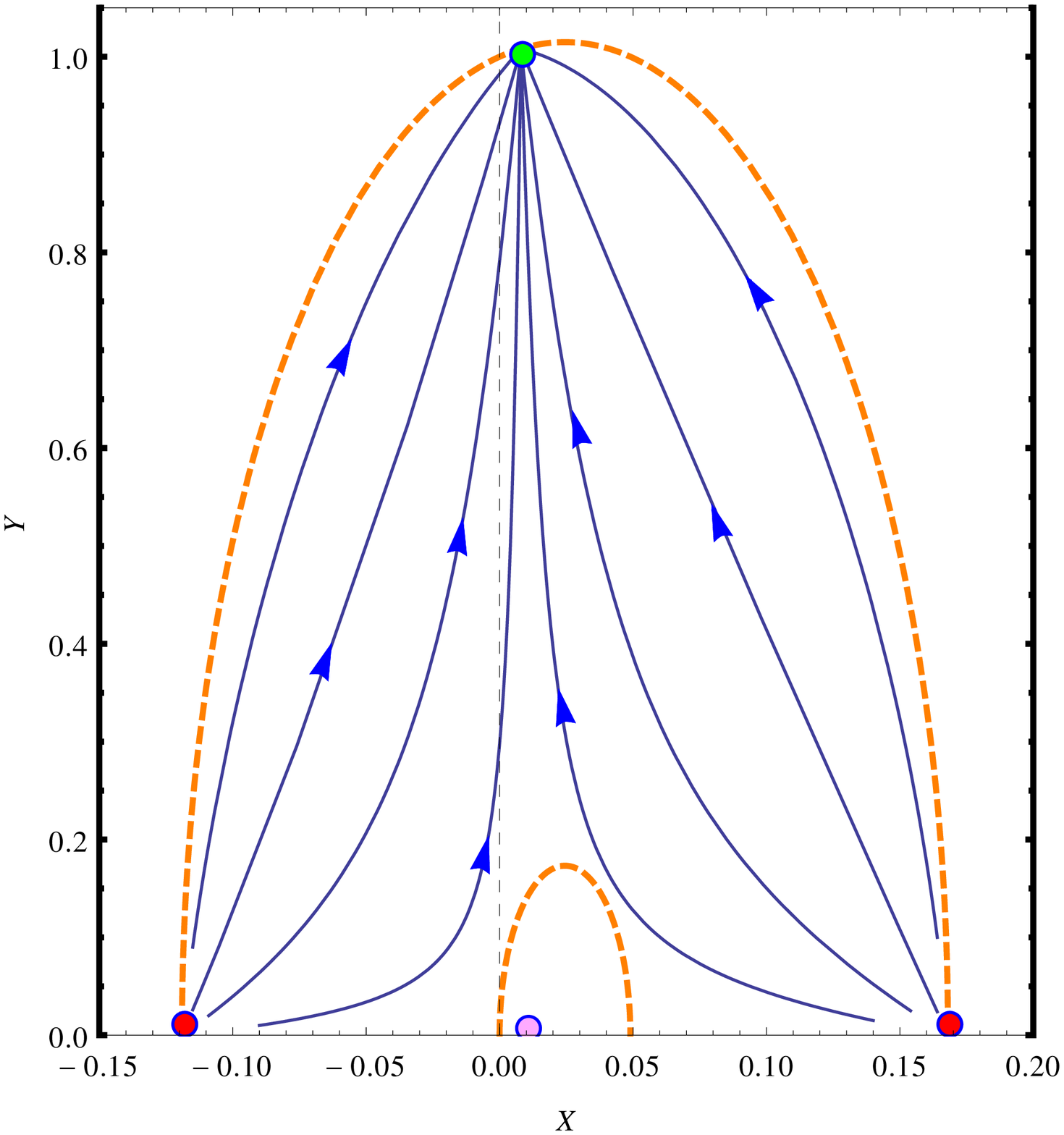}
     \caption{{\footnotesize The ($X-Y$) space}}\label{fig:jordan_xy}
   \end{subfigure}
   \begin{subfigure}{0.45\linewidth} \centering
     \includegraphics[scale=0.27]{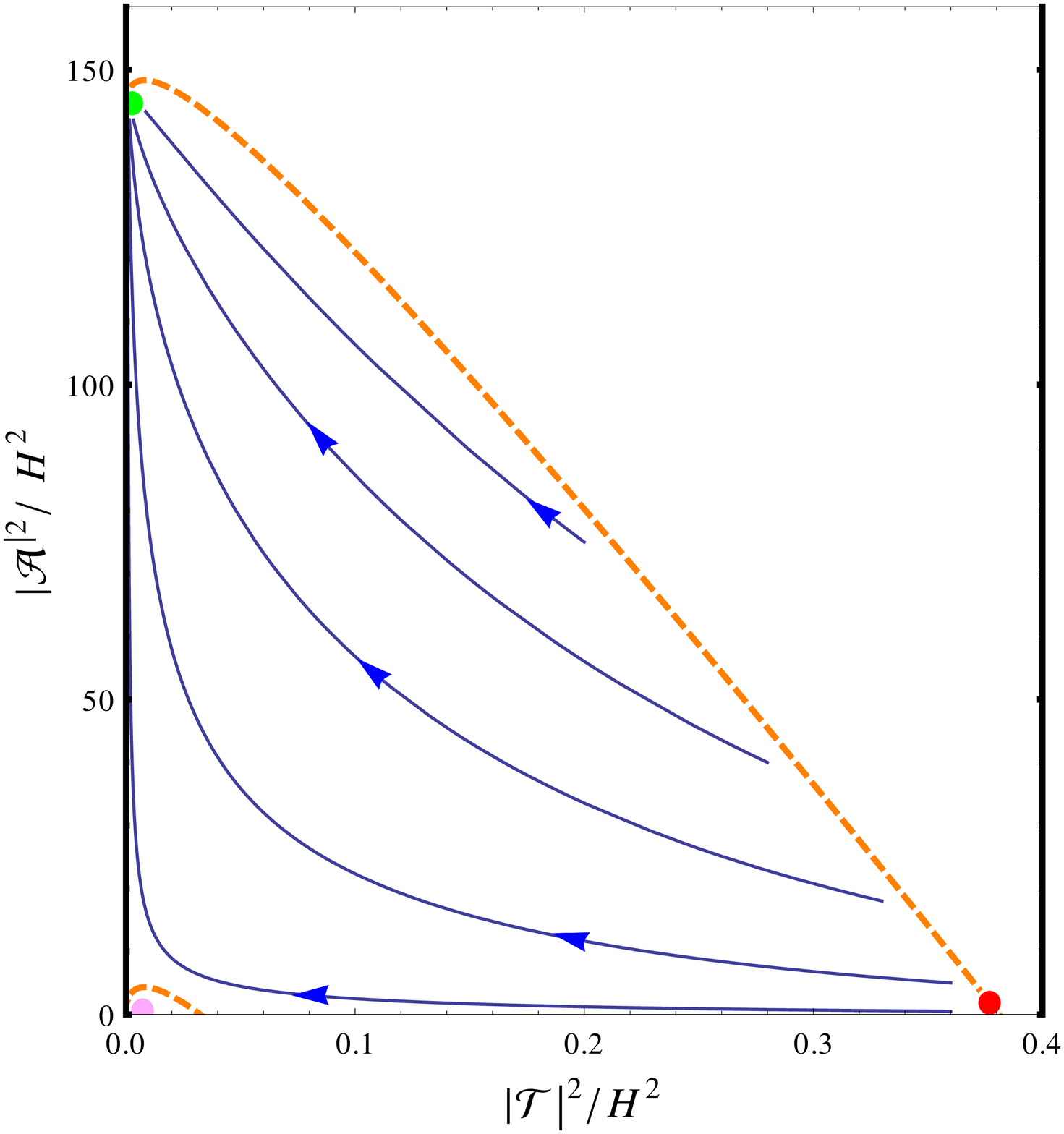}
     \caption{{\footnotesize The (${|\cT|}^2 - {|\cA|}^2$) space}}\label{fig:jordan_ta}
   \end{subfigure}
\caption{{\footnotesize The phase portraits in the two phase planes when $\fw=50$. The dots represent the positions of 
the critical points within the physically allowed region. The arrows mark the direction of evolution of trajectories 
over time. The physically viable regions in both phase planes are bounded by the dashed curves.}} \label{fig:fig4}
\end{figure}

Decomposing the system into matter and dark energy sectors \cite{tsujikawa}, we define the Jordan frame matter and DE density parameters as \footnote{It must be mentioned here that this is not the only recipe of decomposing the system. In \cite{ssasb2} we presented another interpretation for the matter and dark energy sectors of the Jordan frame. But irrespective of how we choose to decompose the system, the field Eqs. (\ref{JordanEom1}) - (\ref{JordanEom4}), which ultimately determine the stability of the system, remain unaltered. As such in this work we adhere to the conventional definitions of the matter and DE sectors as often found in literature related to the Brans-Dicke and other non-minimally coupled scalar-tensor theories of gravity. }

\be
\O_{(m)} ~=~ \fr{\r_m}{3 \F H^2}
\ee
and
\be
\OX ~\equiv~ 1-\O_{(m)} ~=~ \fw X^2 + Y^2 - \sq{6} X
\ee
 Just as in the Einstein frame, all points in the phase plane here do not comply with physically relevant solutions. Since in our model dark energy  and matter account for all the energy present in the universe at any epoch, Eq. (\ref{JordanEom1}) dictates that only those points which satisfy the inequality:

\be \label{Jordan_hubble_constraint}
0 ~\leq~ \OX ~\leq~ 1
\ee
qualify for our analysis. Keeping in mind that our system is again symmetric under the transformation $Y \rightarrow -Y$ and that $\l = 1$ for the specific scalar field potential under consideration 
(Eq.(\ref{BDparameter})), we proceed just as we did in section \ref{sec:field_eqns} to find the critical points for our system of autonomous equations. These critical points and their domains of physical
relevance are listed in Table\ref{Jordan_table_roots}. Also given are the dark energy density parameter and the total E.o.S. parameter for the universe at these respective critical points. The total E.o.S. parameter at any point on the phase plane can be computed using the relation:

\be
\o ~=~ -1 -\fr 2 3 \fr {\dot{H}} {H^2}
\ee
Having found $\OX$ and $\o$, the EoS parameter for the dark energy sector can be computed using the relation:
\be
\oX \OX ~=~ \o
\ee
It now remains to establish the nature of the solutions as dictated by each critical point. Following the same recipe as in the case of Einstein frame, we establish the existence of at most two stable critical points for our model formulated in the Jordan frame. The details of this part have been tabulated in Table\ref{Jordan_table_eigenvalues}. Evolution of some select physical trajectories in the ($X-Y$) and ($|\cT|^2 - |\cA|^2$) phase planes, for the case when $\fw=50$ is presented in Fig. \ref{fig:fig4}.

\begin{table}[h] 
\centering
\renewcommand{\arraystretch}{2}
\begin{tabular}{||c|c|c|c|c|c||}
\hline
\multirow{2}{*}{Name} & \multirow{2}{*}{$X_c$} & \multirow{2}{*}{$Y_c$} & Domain of physical & \multirow{2}{*}{$\O_{DE}$} & \multirow{2}{*}{$\o$}  \\
  &  &  & relevance &  &  \\
\hline\hline
$\#~1$ & $\fr {\sq{3} - \sq{2 \fw +3}} {\sq{2} \fw} $ & 0 & $ \fw \in (-\fr 3 2 , \infty)$ & 1 & $ \fr {6+ 3\fw - 2\sq{6\fw + 9}}{3\fw} $\\
\hline
$\#~2$ & $\fr {\sq{3} + \sq{2 \fw +3}} {\sq{2} \fw} $ & 0 & $ \fw \in (-\fr 3 2 , \infty)-\lbrace 0 \rbrace$ & 1 & $ \fr {6+ 3\fw + 2\sq{6\fw + 9}}{3\o} $ \\
\hline
$\#~3$ & $\fr 1 {\sq{6} \le(1 + \fw \ri)}$ & 0 & $ \fw \in [-\fr 4 3 , -\fr 6 5] $ & $ -\fr {(5\fw + 6)} {6(1 + \fw)^2}$ & $\fr 1 {3(1+ \fw)} $ \\
\hline
$\#~4$ & $\fr 1 {\sq{6} \le(1 + \fw \ri)}$ & $\pm \fr {\sq{ (2\fw + 3) (3\fw + 4)}} {\sq{6} (1+ \fw)}$ & $ \fw \in [-\fr 4 3 , \infty) - \lbrace -1 \rbrace $ & 1 & -1\\
\hline
$\#~5$ & $ - \sq{\fr 3 2}$ & $ \pm \fr {\sq{3\fw + 4}} {\sq{2}} $ & $ \fw \in \lbrace - \fr 4 3\rbrace  $ & $ 3\fw + 5 $& -1\\
\hline
\end{tabular}
\renewcommand{\arraystretch}{1}
\caption{ 
{\footnotesize The critical points for the autonomous system  along with the domain of the parameter $\fw$ for which these critical points exist in the physically relevant region of the phase plane. Also given are the dark energy density parameter, $\OX$ and total E.o.S. parameter for the system, $\o$ for the universe at each of these critical points.}}
\label{Jordan_table_roots}
\end{table}

\begin{table}[h]
\centering
\renewcommand{\arraystretch}{1.8}
\begin{tabular}{||c|c|c|c|c||}
\hline
Name & $\m_1$ & $\m_2$ & Type & Nature \\
\hline\hline
$\#~1$ & $ \fr {3+3\fw-\sq{9+6\fw}}{\fw} $ & $ \fr {3+3\fw-\sq{9+6\fw}}{\fw} $ & Nodal Source $ \forall ~ \fw \in (-\fr 3 2 , \infty)- \lbrace - \fr 4 3 \rbrace$ & Unstable  \\
\hline
\multirow{2}{*}{$\#~2$} & \multirow{2}{*}{$ \fr {3+3\fw+\sq{9+6\fw}}{\fw} $}  & \multirow{2}{*}{$ \fr {3+3\fw+\sq{9+6\fw}}{\fw} $}  & Nodal Sink $\forall ~ \fw \in (- \fr 4 3 , 0)$ & Stable  \\
  &  &  & Nodal Source $\forall ~ \fw \in (- \fr 3 2 , - \fr 4 3) \cup (0, \infty)$ & Unstable\\
\hline
$\#~3$ & $ \fr {(3\fw + 4)}{2(1+ \fw)} $ & $ - \fr {(3\fw + 4)}{2(1+ \fw)} $ & Saddle Point  $ \forall ~ \fw \in (-\fr 4 3 , -\fr 6 5] $  & Unstable    \\
\hline
\multirow{2}{*}{$\#~4$} & \multirow{2}{*}{$ - \fr {(3\fw + 4)}{(1+ \fw)} $}  & \multirow{2}{*}{$ - \fr {(3\fw + 4)}{(1+ \fw)} $}  & Nodal Source $\forall ~ \fw \in (- \fr 4 3 , -1)$ & Unstable  \\
  &  &  & Nodal Sink $\forall ~ \fw \in (-1, \infty)$ & Stable\\
\hline
$\#~5$ & $ \fr {\sq{3} (3\fw + 4)} {\sq{(2\fw +3)}} $ & $ - \fr {\sq{3} (3\fw + 4)} {\sq{(2\fw +3)}} $ & Indeterministic  &  Indeterministic   \\
\hline
\end{tabular}
\renewcommand{\arraystretch}{1}
\caption{{\footnotesize The eigenvalues of matrix $\cM$ corresponding to each critical point along with the type and nature of the critical point in the Jordan frame.} }
\label{Jordan_table_eigenvalues}
\end{table}

Analysing the dynamics of the universe and it's constituent components at each of these five possible critical points, it is revealed that the critical point $ \#~ 4$ is representative of our model in the Jordan frame. Here the Hubble parameter saturates to a constant value asymptotically while the scalar field scales as a power law of the scale factor. Quantitatively at this critical point,

\be
H^2 ~=~ 2 m^2 \fr { (1 + \fw)^2 }{(3 \fw + 4)}
\ee
and
\be
\fr {\F}{\F_0 } ~=~ a^{{(1+ \fw)}^{-1}}
\ee
In view of Eqs.(\ref{BDparameter}) these relations are in exact conformity with our analytical results found in \cite{ssasb2}. As mentioned in Table \ref{Jordan_table_eigenvalues}, the system defined by critical point $ \# ~4$ is stable as long as, 
\be \label{constraint_omega_jordan}
\fw ~>~ -1
\ee
which translates to:
\be \label{constraint_beta_jordan}
\b ~<~ \fr 1 2
\ee
This upper bound on $\b$ happens to be identically same as that obtained from our Einstein frame analysis in section \ref{sec:phase plane analysis},(Eq. (\ref{constraint_beta_einstein})). We have thus successfully established the stability of our analytical results in the Jordan frame as well. 

Let us conclude this work with a brief discussion on what constitutes the physically admissible region of the phase space. As mentioned earlier, this region is constituted by set of points in the phase plane which give rise to a matter density parameter lying anywhere between (or even equal to) zero and unity. In the Jordan frame this condition boils down to selecting points on the phase space which lie on the curves:

\be
\fw X^2 + Y^2 - \sq{6} X ~=~ C
\ee
with 
\be
0 ~\leq~ C ~\leq~ 1
\ee
Clearly this is an equation of a conic section, eccentricity of which is effected by the parameter $\fw $. The curves enclosing the relevant region describe:

\bi
\item An ellipse when $\fw  > 0 $
\item A parabola when $\fw = 0 $ 
\item A hyperbola when $\fw < 0 $
\ei

\begin{figure}[h]
\centering
   \begin{subfigure}{0.45\linewidth} \centering
     \includegraphics[scale=0.8]{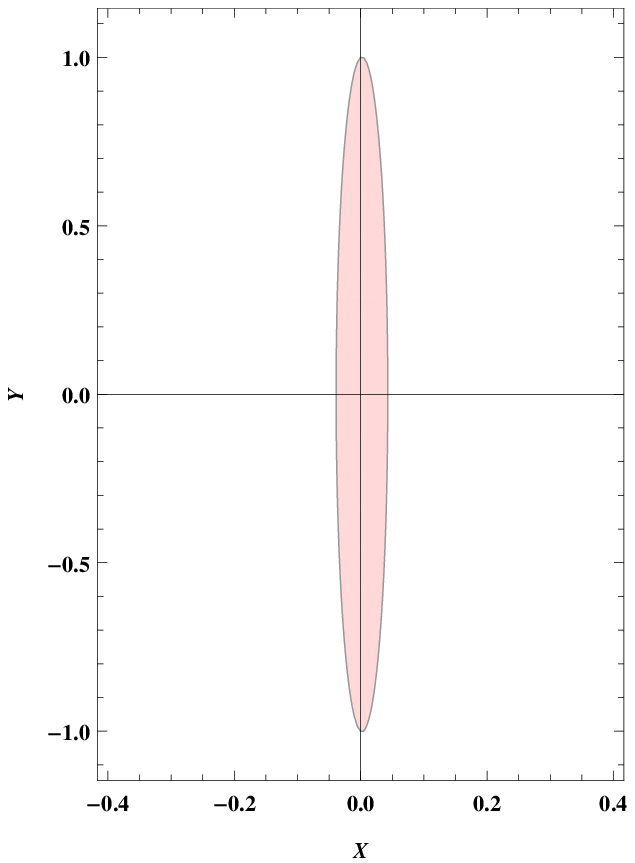}
     \caption{{\footnotesize  $\fw = 600$}}\label{fig:fig5A}
   \end{subfigure}
   \begin{subfigure}{0.45\linewidth} \centering
     \includegraphics[scale=0.8]{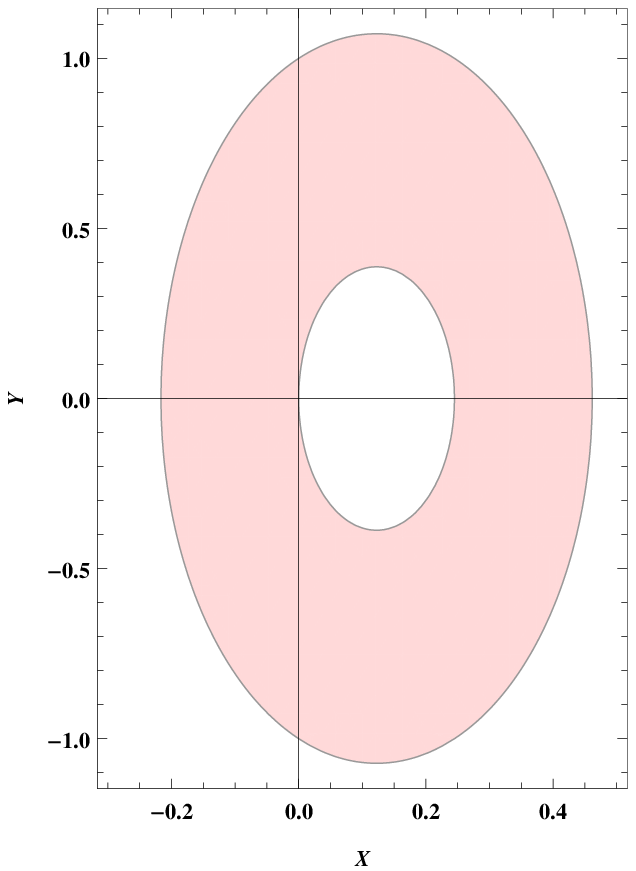}
     \caption{{\footnotesize  $ \fw = 10 $}}\label{fig:fig5B}
   \end{subfigure} 
   \begin{subfigure}{0.45\linewidth} \centering
     \includegraphics[scale=0.8]{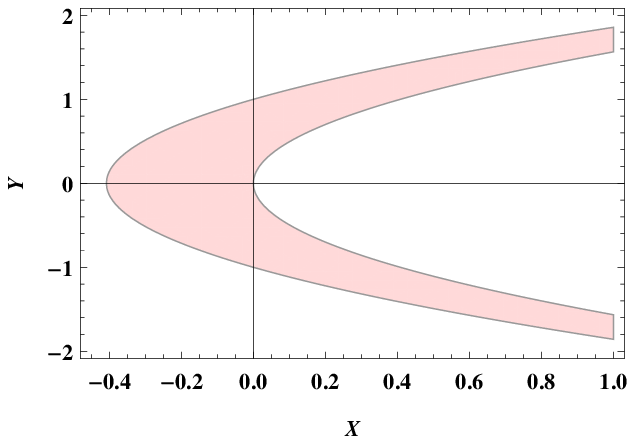}
     \caption{{\footnotesize  $ \fw = 0 $}}\label{fig:fig5C}
   \end{subfigure}
   \begin{subfigure}{0.45\linewidth} \centering
     \includegraphics[scale=0.8]{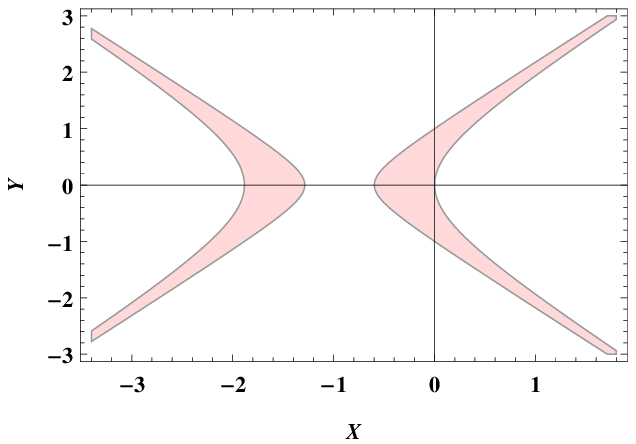}
     \caption{{\footnotesize  $ \fw = -1.3 $}}\label{fig:fig5D}
   \end{subfigure}
\caption{{\footnotesize In shade are the regions of the phase space where $\O_{(m)}$ and $\OX$ exist within the physical bounds. The eccentricities of the curves which bound these regions are determined by the parameter $\fw$. The curves are respectively elliptic, parabolic and hyperbolic in nature when $\fw$ is greater, equal or smaller than zero. }} \label{fig:fig5}
\end{figure}

It is noteworthy that both branches of the hyperbola, although disjoint from each other, play a part in accommodating the critical points and the relevant trajectories for the system. Also worth noting is the fact that these curves are not centred at the origin, but move along the abscissa as
the value of $\fw$ changes. The ellipses in particular undergo a very noticeable change over the domain of the model parameter. In an asymptotic limit when $\fw \rightarrow \infty $, the relevant region is reduced to a segment on the ordinate. The shape of these curves and the region they bound are shown in Fig. \ref{fig:fig5}. for four different values of the parameter, $\fw$. One might wonder why we didn't
encounter any other conic section except the circle (whose centre and radius too remained unaltered) in the Einstein frame over the entire range of the model parameter? The answer lies in Eq.(\ref{action}). 
The fact that we had defined a canonical scalar field using the original scalar field $\F$ (or $\f$) in the Einstein frame took care of this problem. In absence of such a redefinition, we would get an ellipse, parabola or a hyperbola depending on the coefficient of the kinetic term.

\section{Conclusion} \label{sec:conclusion}

Solutions of a set of coupled differential equations have little meaning physically if they do not persist once the system is subjected to perturbations in the solution space. We have in this work established the stable nature of the analytical solutions found in \cite{ssasb2} against any such perturbations,  in Jordan as well as in the Einstein frames. These solutions dictate the dynamics of the universe (and it's constituents) under
the non-minimally coupled MST formalism in the standard cosmological setup. The field equations of the model in either frame can be reduced to an
autonomous set of first order differential equations, which when analysed in an asymptotic limit yield one or more solutions called the \emph{critical points}. The nature of these critical points can then be analysed by applying the linear perturbation theory in the solution space. In Einstein frame the autonomous set of coupled differential equations were found to be very similar to that of the quintessence model in presence of dust. The only
difference was the presence of an extra term in our case whose origin could be traced to the non-minimal coupling between the matter and the scalar field sectors. The effect of this extra term could be clearly seen on the resulting solutions (critical points). The system in Einstein frame allowed
for the existence of $5$ unique critical points when the model parameter, $b \in \le(0, 1 \ri)$. For the cases when $b \in ( 1, \infty)$, this number reduced to $3$. A stable point was shown to exist for our system irrespective of the value of $b$. The solutions to the field
equations at this critical point exactly matched the analytical solutions found in \cite{ssasb2}. 

In contrast to the quintessence model, the saddle point now was no longer fixed at the origin over the domain of the model parameter. Also the critical point which  earlier lead to the scaling/tracking solutions no longer existed now. As a cross check, in the limiting case $b \rightarrow 0$ our results did infact reduce to that of the quintessence model. Another significant difference over the minimally coupled scalar field models was the
region of phase plane which supported physically relevant solutions.  The quintessence model for example allows for the critical points to exist anywhere in the phase plane
within (or on) the boundary of the unit circle centred at the origin. In our case critical points which did not lie on the circumference of the unit circle resulted in a cosmology where the matter and dark energy density parameters eventually crossed the physical bounds rendering the respective solutions un-physical in nature. This was a direct manifestation of the fact that the scalar field used to set up the phase space coordinates remained
non-minimally coupled to the matter sector at all times. For the solutions to remain physically valid, this result stipulated the presence of a matter sector which vanishes completely over time.
Indeed such was the nature of the solutions emanating from the stable critical point, which existed on the aforementioned boundary of the unit circle for all possible values of the model parameter. The stable solutions were shown to give rise to a dynamic dark energy sector and an effective
matter sector which scaled in conformity with the observations. The interplay of the two sectors resulted in a cosmology which could best be described as small fluctuations over a base $\L$CDM picture. These fluctuations, which were quantified in terms of the model parameter $b$, bore their origin to
the dynamic nature of the dark energy sector. To ensure the model remained consistent with the cosmological observations, numerical constraints were placed on the fluctuations and consequently on the value of $b$. Our analysis in the Einstein frame yielded the bound, $b < 1/ \sq{3}$. 

To study the system and it's properties at a finite instant of time, we solved the set of autonomous equations numerically. On top of confirming the existence of the stable point, this exercise allowed us to plot the trajectories that our system took in the phase plane to reach the stable
configuration for different sets of initial conditions. Evolution of cosmological parameters over some of these trajectories (with appropriate initial conditions) again showed a marked resemblance to the $\L$CDM case. The equivalence of the two models was broken as expected by the dynamic dark energy
sector. But the difference was at the very best, small. The system was also analysed in terms of the independent torsion modes (rather their norms). This was made possible due to the fact that in context of the MST model, the torsion trace mode $ \le( \cT_\m \ri) $ is sourced by the kinetic part of the
scalar field whereas the pseudo-trace mode $ \le( \cA_\m \ri) $ could be identified with the scalar field potential. The key result that emerged from this analysis was the indispensable role played by the pseudo-trace mode in driving the system to a stable configuration. It's presence, however
small, ensured the culmination of a phase trajectory at the stable point. Next the system was analysed in the Jordan frame. In the Jordan frame the MST model reduces to a Brans-Dicke model in presence of a massive scalar field potential and dust. The Brans-Dicke parameter ($\fw$)
effectively works as the model parameter. The existence of $5$ critical points and the nature of solutions emanating from them were verified using both analytical and numerical techniques over the range of $\fw$. The stable solutions in this frame were again shown to be in accordance with the solutions obtained in \cite{ssasb2}. Physical compatibility constrained the Brans-Dicke parameter to a domain, $\fw > -1$. This fulfilled our primary objective of verifying the stability of the non-minimally
coupled MST model in presence of dust. We concluded by illustrating the effect of the Brans-Dicke parameter on the region of phase plane deemed physically relevant in the Jordan frame.

Some open questions that naturally arise are: (i) We have successfully established the stability of the MST model in the solution space. But is this model stable under density perturbations as well? (ii) How would the trajectories change if we allow for the radiation sector to exist? (iii) What happens in the torsion phase space if we tweak around with the scalar field potential? Work is in progress to address some of these questions and we hope to share our findings with the community very soon.

\begin{acknowledgements}
The work of ASB was supported by the Council of Scientific and Industrial Research (CSIR), Government of India. SS acknowledges the R \& D Grant DRCH/R \& D/2013-14/4155, Research Council, University of Delhi.
\end{acknowledgements}




  
  \renewcommand{\theequation}{A-\arabic{equation}}
  \setcounter{equation}{0}  

\section*{Appendix} \la{sec:auton_sys}

Having constructed an autonomous system of equations representing the dynamics of a system in the phase space, we discuss here the theory behind finding the critical points and consequently the stability of solutions they yield. Although for brevity we restrict ourselves to a two dimensional phase space, these results can easily be extended to higher dimensions as well. Let us represent autonomous equations as:

\be \label{f,g eqn}
\fr {dX}{dN} ~=~ f(X,Y) ~~~~~,~~~~~ \fr {dY}{dN} ~=~ g(X,Y)
\ee
A \emph{critical point} or a \emph{fixed point}, $(X_c,Y_c)$, is defined as any point in the phase plane  that satisfies the conditions:

\be
f~|_{\le(X_c, Y_c \ri)} ~=~ 0 ~~~~~,~~~~~ g~|_{\le(X_c, Y_c \ri)} ~=~ 0 
\ee
A critical point serves as an equilibrium solution to the set of autonomous differential equations. To investigate the nature of a critical point, consider small perturbations $\le( \d X, \d Y \ri)$ such that:

\be
X = X_c + \d X ~~~,~~~ Y = Y_c + \d Y
\ee
In view of these perturbations, Eqs.(\ref{f,g eqn}) reduce to:
\be
\fr {d}{dN}~
   \begin{pmatrix}
        \d X \\
        \d Y
     \end{pmatrix}
 =\cM \begin{pmatrix}
      \d X \\
      \d Y
    \end{pmatrix}
\ee    
where
\begingroup                                    
\renewcommand*{\arraystretch}{1.5}
\be \label{matrix}
\cM \equiv 
    \begin{pmatrix}
        \fr{\pa f}{\pa X} & \fr {\pa f}{\pa Y} \\
        \fr{\pa g}{\pa X} & \fr {\pa g}{\pa Y}
     \end{pmatrix}_{\le(X=X_c , Y=Y_c \ri)}
\ee
\endgroup
The evolution of these linear perturbations around a critical point can be written in terms of the two eigenvalues 
for the matrix $\cM$ at that point, say $\le(\m_1,\m_2 \ri)$ as:

\bea
\d X ~=~ C_1 e^{\m_1 N} + C_2 e^{\m_2 N} \\
\d Y ~=~ C_3 e^{\m_1 N} + C_4 e^{\m_2 N}
\eea
Here $C_i$ are arbitrary constants. If the perturbations die over time, i.e. $\le(\d X, \d Y \ri)\longrightarrow 0$ as $N \longrightarrow \infty$, the critical 
point in question acts as an attractor/stable point. It is thus the sign of the eigenvalues $\le(\m_1,\m_2 \ri)$ which determines 
the stability of a critical point. The different possible types of eigenvalues that can exist and the nature of the critical point 
they result in can be summarized as:

\ben
\item \textbf{Real eigenvalues of the same sign}: Depending on the sign of the eigenvalues, two cases arise,
\bi
\item Both eigenvalues $\m_1, \m_2 < 0$ : In this case any perturbations in coordinates around the critical point 
tend to die out over time, hence leading all trajectories in the vicinity to terminate at the critical point. Due to 
the nature of phase portraits around such a critical point, it is called an \emph{attractor} or a \emph{nodal sink}. 
The nature of the attractor is asymptotically stable.

\item Both eigenvalues $\m_1, \m_2 > 0$ : The behaviour of a critical point under this condition is exactly opposite 
to that of an attractor. The perturbations in this case build up over time taking trajectories away from the 
critical point. The phase diagram portrays a picture of phase lines being repelled about this point. Hence such a critical point 
is termed a \emph{nodal source} and is unstable in nature.
\ei

\item \textbf{Real eigenvalues of opposite sign}: Such a combination of eigenvalues result in a \emph{saddle point}, 
a critical point that acts as an attractor along one particular direction in the phase plane (the attractor axis) and 
as an unstable point along the direction normal to the attractor axis. Trajectories might tend towards this point but are eventually repelled away. The only trajectories that manage to reach this critical point are those which begin from point on the 
attractor axis and proceed along it. The saddle point acts as an attractor for such trajectories. A saddle point is otherwise unstable in nature.

\item \textbf{Complex eigenvalues}: Complex eigenvalues lead to a spiral behaviour of the trajectories. Whether the 
trajectories recede or approach the critical point depends on the sign of the real part of the eigenvalues,

\bi

\item $Re~\{\m_1, \m_2\} < 0$ : The trajectories in this case spiral towards the critical point lending an asymptotically 
stable nature to it. Such a point is referred to as \emph{spiral sink}.

\item $Re~\{\m_1, \m_2\} > 0$ : In this case, the trajectories spiral away from the critical point leading to an unstable nature. 
Such a point is called a \emph{spiral source}.
\ei

\item \textbf{Pure imaginary eigenvalues}: The trajectories in this case describe a circle, or more generally an ellipse 
with the critical point at it's center. The critical point in question is hence appropriately termed a \emph{center}. 
Such a point is stable in nature.

\item \textbf{Vanishing eigenvalues}: If the eigenvalues $\m_1 = \m_2 = 0$, then the linear stability theory breaks down and one has to resort to 
alternative theories to uncover the nature of such a critical point.

\een



\end{document}